\documentclass[12pt,draftclsnofoot,onecolumn]{IEEEtran}
 \usepackage[english]{babel}
 \usepackage{fancyhdr}
\usepackage{epsfig}
\usepackage{threeparttable}
\usepackage{epsf,epsfig}
\usepackage{amsthm}
\usepackage{amsmath}
\usepackage{amssymb}
\usepackage{amsfonts}
\usepackage[noadjust]{cite}
\usepackage{dsfont}
\usepackage{graphicx,amssymb,amsmath}
\usepackage{subfigure}
\usepackage{xcolor}
\usepackage{soul}
\usepackage{algorithmic,algorithm}
\usepackage{url}
\usepackage{bm}
\usepackage{mathtools}

\usepackage{newtxmath}
\newtheorem{theorem}{Theorem}

\newtheorem{lemma}{Lemma}

\newtheorem{example}{\bf Example}
\newtheorem{remark}{\bf Remark}
\newtheorem{granularity_constraint}[theorem]{Granularity Constraint}

\def\proof{\noindent{\emph{Proof:} }}

\def\phi{\varphi}

\def\({\left(}
\def\){\right)}

\def\b0{{\mathbf{0}}}

\begin{document}
\graphicspath{{figure/}}

\title{\huge Adaptive Subcarrier, Parameter, and Power Allocation for Partitioned Edge Learning Over Broadband Channels}
\author{Dingzhu Wen, Ki-Jun Jeon, Mehdi Bennis, and Kaibin Huang     \thanks{\setlength{\baselineskip}{13pt} \noindent D. Wen and K. Huang are with  The  University of  Hong Kong, Hong Kong. K.-J. Jeon is with the LG Electronics, Korea. M. Bennis is with University of Oulu, Finland.  Corresponding email: huangkb@eee.hku.hk. }
}
\maketitle

\begin{abstract}
In this paper, we consider  \emph{partitioned edge learning} (PARTEL), which implements parameter-server training, a well known distributed learning method, in a wireless network. Thereby, PARTEL leverages distributed computation resources at edge devices to train  a large-scale \emph{artificial intelligence} (AI) model by dynamically partitioning the model into  parametric blocks  for separated updating at devices. Targeting broadband channels, we consider the  joint control  of  parameter allocation,  sub-channel allocation, and transmission power  to improve the performance of PARTEL. Specifically, the policies for joint SUbcarrier, Parameter, and POweR allocaTion (SUPPORT) are optimized under the criterion of minimum learning latency. Two cases are considered. First, for the case of decomposable models (e.g., logistic regression), the latency-minimization problem is a mixed-integer program and non-convex. Due to its intractability, we develop a practical solution by integer relaxation and transforming it into an equivalent convex  problem of  model size maximization under a latency constraint. Thereby, a low-complexity algorithm is designed to compute the SUPPORT policy. Second, consider the case of \emph{deep neural network} (DNN) models which can be trained using PARTEL by introducing some auxiliary variables. This, however, introduces constraints on model partitioning reducing the granularity of parameter allocation. The preceding policy is extended to DNN models by applying the proposed techniques of load rounding and proportional adjustment to rein in latency expansion caused by the load granularity constraints. %Finally, experiments using real data show that joint SUPPORT can substantially reduce the latency of PARTEL for decomposable models (e.g., 31\%) and DNN models (e.g., 42\%). 
\end{abstract}

%\hl{Please don't use "vspace" to adjust space. It is not professional}

%\hl{The below caption space is usually large. Something is wrong with your settings. Please check.}
\section{Introduction}
Edge machine learning is an  area concerning the deployment of  learning algorithms at the network edge to gain  low-latency access to data and computation resources distributed at a large number of edge devices \cite{park2020communication}.  In this work, we study the efficient implementation of the well-known method of parameter-server training \cite{li2013parameter} in a broadband  system (e.g., 3GPP 5G) to exploit  distributed computation resources at many devices to scale up model training. To this end,  several key operations, namely computation-load allocation (via  model partitioning), sub-channel allocation, and power control, are jointly designed under the criterion of minimum learning latency.

Two main methods for distributed learning are federated learning \cite{mcmahan2016communication,konevcny2016federated} and parameter-server training \cite{li2013parameter, carreira2014distributed,choromanska2019beyond}, which are designed  for different scenarios and features. The key feature of federated learning is its preservation of data privacy. Based on distributed  implementation of  \emph{stochastic gradient descent} (SGD),  federated learning iterates the separate training of a downloaded model at multiple devices using their local data,  and the uploading and aggregation of local models (or local stochastic gradients) to yield a more accurate global model  \cite{mcmahan2016communication,konevcny2016federated}. The avoidance of direct data uploading protects their privacy. Though it is similar to federated learning in implementing distributed SGD, the parameter-server training, which is of our interest,  has one  distinction. Its purpose is to scale up learning using many resource-constrained machines in a closed network where data privacy is not a concern \cite{li2013parameter}. To this end, the model is partitioned to allow each device to train only a part of the model instead of the whole as in federated learning. This overcomes the resource constraints of devices and reduces their energy consumption. Moreover, training data are downloaded from a server to devices at the beginning of each round, avoiding their need of  persistent storage space.

A current   main theme in the field  of edge learning is the  design of wireless techniques to support efficient deployment of federated learning, resulting in an area called \emph{federated edge learning} (FEEL) \cite{zhu2020toward}. The effort on overcoming the communication bottleneck of FEEL has led to the design of  a new class of multi-access techniques realizing over-the-air model aggregation\cite{zhu2019broadband, amiri2020machine, ShiyuanmingAirComp, du2020high, jeong2018communication} and \emph{radio resource management} (RRM) techniques \cite{yang2019scheduling, ren2020scheduling, chen2019joint, shi2019device, ren2019accelerating}. Moreover, researchers have designed energy efficient RRM techniques to tackle the challenge of executing a complex  learning task at energy constrained devices in a FEEL system\cite{yang2019energy, zeng2020energy, mo2020energy}.  Recently, researchers have also explored the efficient implementation of parameter-server training over wireless channels, resulting in a framework called \emph{partitioned edge learning} (PARTEL) \cite{wen2020joint}. Let  \emph{parameter allocation} refers to the system operation that to balance computation loads,  the server divides the model into  parametric blocks of variable lengths and  allocate them to devices for separate training. To reduce the learning  latency, the technique of joint parameter allocation  and resource allocation is proposed in \cite{wen2020joint}, which jointly adapts parameter and bandwidth allocation   to devices'  channel states and computation capacities. For simplicity, the prior work assumes narrowband channels, for which the management of uplink radio resource reduces to bandwidth allocation.  In this work, we design  low-latency PARTEL techniques for a broadband system (e.g., 3GPP 5G) with frequency selective channels. In this case, the frequency resource is managed via sub-channel allocation, which is much more complex than bandwidth allocation. The complexity arises from the fact that the sub-channels of each device have different gains and devices have different channel realizations. Consequently, even if  the allocated bandwidths  are fixed, reshuffling the assignments of sub-channels varies devices' communication rates. Therefore,  jointly designing sub-channel and parameter allocation poses a new challenge that cannot be tackled using  the solution in \cite{wen2020joint}.

The optimal sub-channel allocation is well known to be an integer optimization problem that is NP hard \cite{wong1999multiuser, ng2012energy, jang2003transmit}. For the  conventional multiuser communication systems, the classic approximate-solution approach has been established in a series of work for the purpose of minimizing  sum power under users' rate constraints \cite{wong1999multiuser, ng2012energy, jang2003transmit}. The essential idea is to  relax the integer program and discover the embedded convexity in the relaxed problem to design a practical algorithm \cite{wong1999multiuser}. In this work, we build on the classic approach to develop a new  solution for the problem of latency minimization in a  broadband PARTEL system. The distinction of the current work is the pursuit of a communication-learning integration approach so as to minmize learning latency in the context of PAETEL. To this end, we jointly design parameter and sub-channel allocations. The considerations of synchronized updates by devices, which is a requirement for parameter-server training \cite{li2013parameter}, and devices' heterogeneous computation capacities introduce more challenges. Existing designs that aim at  generic radio access  cannot tackle the new challenges, which motivate the current work. 

It is also worth mentioning that we also consider a more complex model based on a \emph{deep neural network} (DNN) besides the decomposable model as in \cite{wen2020joint}.  Unlike the latter, the former  is not directly decomposable and requires the modification of learning algorithm to support PARTEL. This introduces additional complexity to the current design. 

By  tackling the above challenges, we design a set of algorithms for joint SUbcarrier, Parameter, POweR allocaTion (SUPPORT), termed  joint SUPPORT. The main contributions of this  work are summarized as follows. 

\begin{itemize}
\item {\bf Joint SUPPORT for Decomposable Models}:  Consider the case of a decomposable model. The problem of latency minimization by joint SUPPORT is an integer program and intractable. A practical solution approach is developed using two techniques. The first is a relaxation of binary subcarrier assignment decisions. The second is the transformation of the relaxed problem into a convex problem of model size maximization under a latency constraint, which is nested in a simple search for the target model size. Considering the convex problem, the properties of three optimal resource-management operations are analyzed and then applied to design an efficient algorithm for computing the desired SUPPORT policy. Via analysis, it is found that the optimal number of parameters assigned to a worker for updating avoids high power consumption due to overloading. For this reason, the  optimal number is derived to be a concave function of its speed and a monotonic decreasing function of its computation power factor. On the other hand, the optimal subcarrier assignment and power allocation over assigned subcarriers are found to favor high channel gains.% and are independent of devices' computation capacities.

\item {\bf Joint SUPPORT for DNN Models}:  Consider the case of a DNN model. The optimization problems for joint SUPPORT in both mini-rounds are shown to have the same form as that in the preceding case except for additional load granularity constraints. This allows the extension of the joint SUPPORT policy for the case of decomposable models to the current case by rounding  down the obtained loads to meet the granularity constraints. Furthermore, the remaining parameters due to rounding are allocated over devices and subcarriers \emph{proportionally} with their rounded loads, thereby reining in the latency expansion caused by the additional constraints.

\end{itemize}
The performance gain of the above algorithms and the findings are corroborated using experiments with a real dataset.

The reminder of the paper is organized as follows. In Section II, the system model is introduced. In Section III, the total learning latency minimization problem is formulated. In Section IV and V, the joint SUPPORT designs are proposed for decomposable models and DNN models, respectively. Section VI presents the experimental results followed by concluding remarks in Section VII.

\begin{figure}[t]
    \centering
    \subfigure[PARTEL system]{\includegraphics[width=0.5\textwidth]{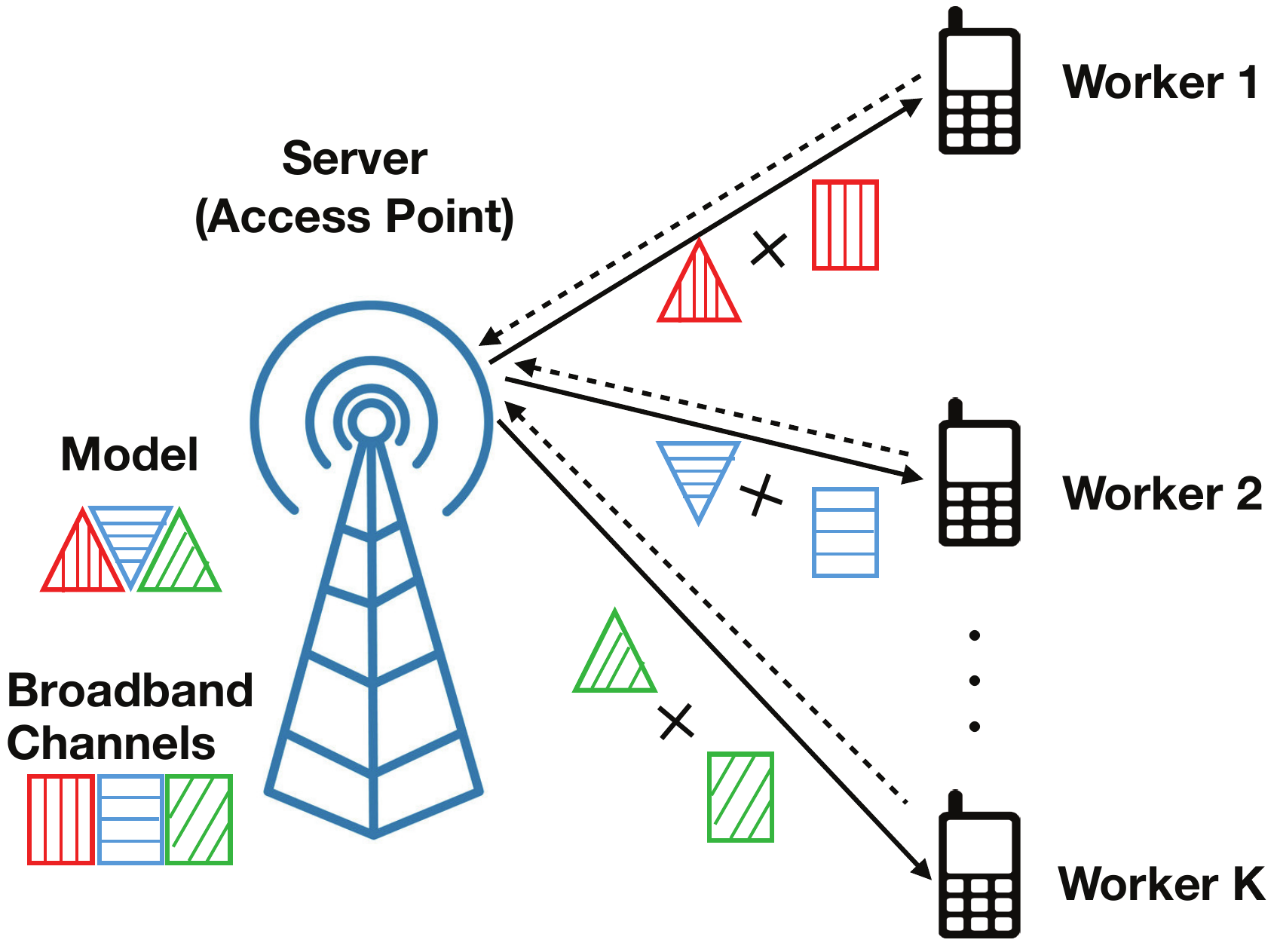}}
    \subfigure[PARTEL operations and their latencies]{\includegraphics[width=0.5\textwidth]{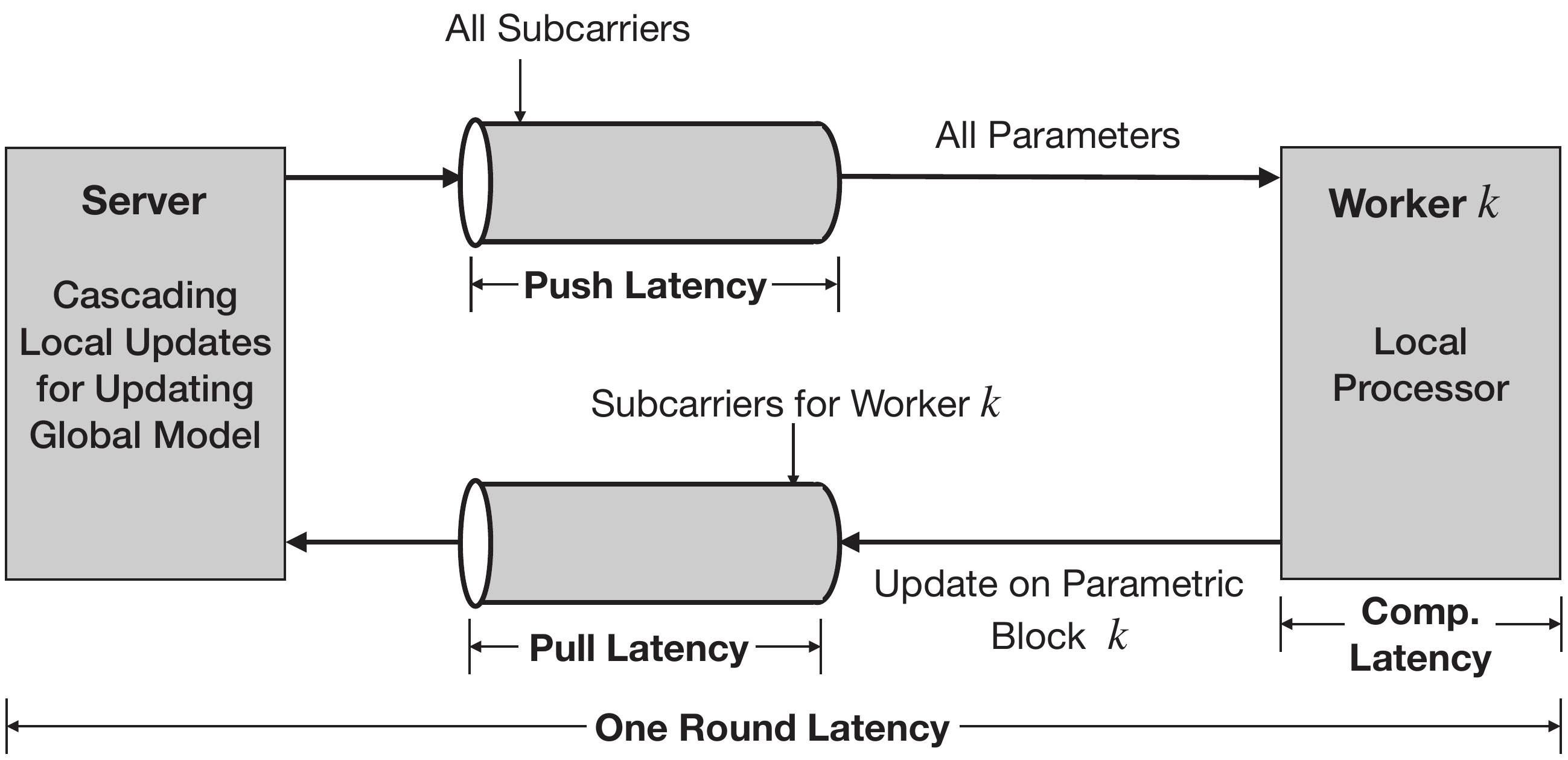}}
    \caption{PARTEL system, operations, and latencies.}\label{fig:SystemModel}
\end{figure}

\section{Models and Metrics}

\subsection{System  Model}

A single cell OFDM system is considered, as illustrated in Fig. \ref{fig:SystemModel}(a). In the cell, there are one server equipped with a single-antenna \emph{access point} (AP) and $K$ single-antenna edge devices, serving as workers. Each worker performs one task assigned by the server.  The server is connected to workers via wireless links. The system bandwidth is divided into $N$ orthogonal subcarriers. The bandwidth of each subcarrier is denoted as $B$. The frequency-selective fading channel is considered, where different subcarriers will experience different channel gains. We assume that the AP has the \emph{channel state information} (CSI) of all links that are useful for subcarrier allocation.  Besides, the channel gains are assumed to be static in one training iteration but vary over different iterations. The uplink channel gain of worker $k$ on the subcarrier $n$ is denoted as $h_{k,n}$. We denote $\{C_{k,n}\}$ as the subcarrier allocation indicators. If the $n$-th subscriber is allocated to worker $k$, then $C_{k,n}=1$. Otherwise, $C_{k,n}=0$.

\subsection{Learning Models}\label{sect:WDLA}

\subsubsection{Decomposable Models}
The large-scale learning tasks with decomposable objective functions (such as logistic regression) can be directly implemented using PARTEL based on the method of block coordinate descent.  According to the literature (e.g., \cite{wen2020joint}), a decomposable objective function can be written as
\begin{equation}\label{eq:ObjLoss}
\mathcal{L}({\bf w})  = \mathcal{F}({\bf w}) + \mathcal{R}({\bf w}),
\end{equation}
where ${\bf w}=\{w_1, w_2, ..., w_L\}^T$ is the parameter vector of the learning model, $L$ is the size of ${\bf w}$, $\mathcal{F}({\bf w})$ is the loss function, and $\mathcal{R}({\bf w})$ is the regularized function (e.g., $L_1$ regularization used to increase sparsity and $L_2$ regularization used to reduce overfitting). Specifically, the loss function can be written as
%\begin{equation}
$\mathcal{F}({\bf w}) = \dfrac{1}{M} \sum\nolimits_{m=1}^M \left| y_m - \phi({\bf w} ; {\bf x}_m) \right|^2$,
%\end{equation}
where $M$ is the size of the dataset, $\{ {\bf x}_m, y_m\}$ is the $m$-th data sample, and $\phi(\cdot)$ is a smooth inference function. The regularized function is a block-separable function, given as
%\begin{equation}
$\mathcal{R}({\bf w}) = \sum\nolimits_{i=1}^{L} \psi(w_i)$,
%\end{equation}
where $w_{i}$ is the $i$-th element of ${\bf w}$ and $\psi(\cdot)$ is the norm (e.g., $L_1$ or $L_2$ norm). During the training, the smoothness of $\mathcal{R}(\cdot)$ decides the method to update the learning model. If $\mathcal{R}(\cdot)$ is smooth, gradient descent algorithm is used. Otherwise, another method called proximal gradient descent, is used.

\subsubsection{DNN Models}
DNN models cannot be directly implemented using PARTEL, as the nested layers therein make the gradient elements of different layers dependent. To make PARTEL feasible and efficient for DNNs, in the sequel, the method of auxiliary variables is used to decompose the DNN models into many independent subproblems \cite{carreira2014distributed,choromanska2019beyond}. 

First, consider a DNN model with $G$ hidden layers. The model parameter matrix is denoted as ${\bf W}$ with the size of $L$ parameters. For an arbitrary layer therein, say layer $g$, the  parameter matrix is denoted as ${\bf W}_g$, the number of neurons is denoted as $I_g$, and the $i$-th neuron parametric vector is denoted as ${\bf w}_{g,i}$. Thereby, the objective function of the DNN model is given by 
\begin{equation}\label{Eq:DNN}
\begin{aligned}
&\mathcal{L}({\bf W}) = \sum\limits_{m=1}^M \left| y_{m} - \mathcal{F} \left({\bf x}_{m} ;   {\bf W} \right)  \right|^2, \\
\text{with }& \mathcal{F} \left({\bf x} ;   {\bf W} \right) =  f_{G+1} \left( ...{\bf f}_2\left( {\bf f}_1\left(  {\bf x} ;  {\bf W}_1  \right) ; {\bf W}_2 \right),...; {\bf W}_{G+1} \right),
\end{aligned}
\end{equation}
where the model parameter matrix can be expressed as ${\bf W} = \left[{\bf W}_1, {\bf W}_2,...,{\bf W}_G,{\bf W}_{G+1} \right]$, the parameter matrix of the $g$-th layer can be expressed as ${\bf W}_g = [{\bf w}_{g,1},{\bf w}_{g,2},...,{\bf w}_{g,I_g}]$,  and ${\bf f}_g \left( {\bf x}_g ;  {\bf W}_g \right)$ is the set of output (activation) functions of the $g$-th layer. %Note that there are two kinds of layers in a DNN, say convolutional layers and fully-connected layers.

%a DNN with $G$ hidden layers has the following objective function: 
%\begin{equation}\label{Eq:DNN}
%\begin{aligned}
%&\mathcal{L}({\bf W}) = \sum\limits_{m=1}^M \left| y_{m} - f \left({\bf x}_{m} ;   {\bf W} \right)  \right|^2, \\
%\text{with }& f \left({\bf x} ;   {\bf W} \right) =  f_{G+1} \left( ...{\bf f}_2\left( {\bf f}_1\left(  {\bf x} ;  {\bf W}_1  \right) ; {\bf W}_2 \right),...; {\bf W}_{G+1} \right),
%\end{aligned}
%\end{equation}
%where ${\bf W} = \left[{\bf W}_1, {\bf W}_2,...,{\bf W}_G,{\bf W}_{G+1} \right]$ is the model parameter matrix, ${\bf W}_g = [{\bf w}_{g,1},{\bf w}_{g,2},...,{\bf w}_{g,I_g}]$ is the model parameter matrix of the $g$-th layer, ${\bf w}_{g,i}$ is the model-parameter vector of the $i$-th neuron in layer $g$, $I_g$ is the number of neurons in layer $g$, and ${\bf f}_g \left( {\bf x} ;  {\bf W}_g \right)$ is the set of output functions of the $g$-th layer.  Note that there are two kinds of layers in a DNN, say convolutional layers and fully-connected layers.

{\bf Auxiliary Variables}:  The method of auxiliary variables is used by introducing one auxiliary variable per neuron per data sample: 
%\begin{equation}
$z_{g,i,m} = f({\bf w}_{g,i}; {\bf z}_{g-1,m}), \; \forall (g,i,m)$,
%\end{equation}
where $f(\cdot)$ is the activation function, ${\bf w}_{g,i}$ is the $i$-th neuron parametric vector in layer $g$,  $z_{g,i,m}$ is the auxiliary variable introduced for the $i$-th neuron in layer $g$ regarding data sample $m$, ${\bf z}_{g-1,m} = [ z_{g-1,1,m},  z_{g-1,2,m}, ..., z_{g-1,I_{g-1},m}]^T$ is the auxiliary variable vector for the layer $(g-1)$ regarding data sample $m$, and $I_{g-1}$ is the number of neurons in the $(g-1)$-th layer. For an arbitrary data sample, say the $m$-th, the corresponding auxiliary matrix for the whole model is denoted as ${\bf Z}_m = [{\bf z}_{1,m},..., {\bf z}_{g,m},...,{\bf z}_{G,m}]$, called \emph{per-sample auxiliary matrix}. Then the overall auxiliary matrix for all samples are denoted as ${\bf Z} = [{\bf Z}_1,...,{\bf Z}_m,...{\bf Z}_M]$.

{\bf Decomposed Optimization}: Following \cite{carreira2014distributed,choromanska2019beyond}, by using the quadratic-penalty method, %in \cite{nocedal2006numerical},  
the problem in \eqref{Eq:DNN} is equivalent to minimizing
\begin{equation}\label{Eq:MAVP1}
\mathcal{L}_{\rm Q}({\bf W};{\bf Z};\mu) =  \sum\limits_{m=1}^M \left|  y_{m} - f_{G+1} \left({\bf z}_{G,m};   {\bf W}_{G+1} \right)  \right|^2+ \mu \sum\limits_{g=1}^G \sum\limits_{m=1}^M  \left\|   {\bf z}_{g,m} - {\bf f}_g({\bf z}_{g-1,m}; {\bf W}_g) \right\|^2,
\end{equation}
where ${\bf z}_{0,m} = {\bf x}_m$ and  $\mu\to+\infty$. In \eqref{Eq:MAVP1}, the nested structure among layers is decoupled. Consequently, the gradients of any two parameters (or auxiliary variables) are independent.

Finally, the problem in \eqref{Eq:MAVP1} can be solved using the alternating optimization over ${\bf W}$ and ${\bf Z}$, i.e., sequentially solving the ${\bf W}$-stage and ${\bf Z}$-stage, defined below, in each training iteration.
\begin{itemize}
\item ${\bf W}$-stage: Fixing the values of ${\bf Z}$, solve the problem of $\min\nolimits_{\bf W} \;\mathcal{L}_{\rm Q}({\bf W};{\bf Z};\mu) $, in which the problem of each neuron is independent and can be written as
\begin{equation}\label{Eq:WStep}
\min\limits_{{\bf w}_{g,i}}\; \sum\limits_{m=1}^M \left| z_{g,i,m} - f({\bf w}_{g,i}; {\bf z}_{g-1,m}) \right|^2,\quad \forall (g,i),
\end{equation}
where ${\bf w}_{g,i}$ and $z_{g,i,m}$ are the parameteric vector and auxiliary variable of  the $i$-th neuron in the $g$-th layer, respectively, ${\bf z}_{g-1,m}$ is the auxiliary variable vector of the $(g-1)$-th layer. Note that one device is allocated a task of updating one or more neuron parametric vectors by solving the subproblems in \eqref{Eq:WStep}.

%the load (parameters) are allocated in terms of hundreds of neuron vectors (equivalently solving hundreds of subproblems).

\item ${\bf Z}$-stage: Conditioned on the values of  ${\bf W}$, solve the problem of $\min\nolimits_{\bf Z} \;\mathcal{L}_{\rm Q}({\bf W};{\bf Z};\mu) $, where the problem of optimizing each per-sample auxiliary matrix is independent of others and is given as
\begin{equation}\label{Eq:ZStep}
\min\limits_{{\bf Z}_m}\; \left| y_m - f_{G+1} ({\bf W}_{G+1}; {\bf z}_{G,m} )  \right|^2  + \mu \sum\limits_{g=1}^G \left\| {\bf z}_{g,m} - {\bf f}_g ({\bf W}_g; {\bf z}_{g-1,m} )  \right\|^2,\quad \forall m,
\end{equation}
where ${\bf Z}_m$ is the per-sample auxiliary matrix corresponding to  data sample $m$. The size of the per-sample auxiliary matrix is $\sum_{g=1}^G I_g$ with $I_g$ being the number of neurons in layer $g$. Note that one device is allocated a task of updating one or more per-sample auxiliary matrices by solving the subproblems in \eqref{Eq:ZStep}.%the load (auxiliary variables) are allocated in terms of hundreds of per-sample auxiliary matrices (equivalently solving hundreds of subproblems).
\end{itemize}

\subsection{PARTEL Architecture}\label{sect:WDLA}

Consider the PARTEL system and operations in Fig. \ref{fig:SystemModel}, that are elaborated as follows.

\subsubsection{Decomposable Models}
The model-parameter vector is partitioned into $K$ disjoint parametric blocks, as ${\bf w} = \{{\bf w}_1,...,{\bf w}_k,...,{\bf w}_K\}$, where ${\bf w}_k$ is allocated to worker $k$ for update, using a downloaded global dataset from the server\footnote{The joint SUPPORT design of this paper can be easily extended to the case of partitioned dataset with multiple groups of workers (each with a data subset). Each group cooperatively updates a same block. The proposed joint SUPPORT can be applied in a hierarchical manner: applied for inter-group resource management and also applied for intra-group management.}. The communication overhead for the server to broadcast the dataset is ignored, as its large power and bandwidth are used and the dataset broadcasting requires only once before the model training. One main benefit of PARTEL is low learning latency, as each resource-constrained worker is required to calculate and transmit the gradient or proximal gradient of only a parametric block instead of the whole parameter vector during each iteration \cite{wen2020joint}. 

In the PARTEL framework, one training iteration of the decomposable models is called \emph{one (communication) round}. As shown in Fig. \ref{fig:SystemModel}(b), there are three phases in each round, as follows.
\begin{itemize}
\item \emph{Push Phase}: The server broadcasts the whole model-parameter vector, ${\bf w}$, to all workers.
\item \emph{Computation Phase}: Each worker computes the update (e.g., gradients or proximal gradients) of its allocated parametric block.
\item \emph{Pull Phase}: All workers upload the updates of their corresponding parametric blocks to the server. The server updates the whole parameter vector. %with gradient descent for $L_2$-regularized objective function (or proximal gradient descent for $L_1$-regularized objective function).
\end{itemize}
The training process in Fig. \ref{fig:SystemModel}(b) iterates when all parametric blocks  are updated in the round, i.e., the tasks of all workers are synchronized in each round.

\subsubsection{DNN Models}
As mentioned, each round of DNN models comprises two stages: ${\bf W}$-stage and ${\bf Z}$-stage, described as follows.
\begin{itemize}
\item ${\bf W}$-stage: The parameter matrix  ${\bf W}$ is divided into $K$ blocks, with each being updated by one worker. To avoid inter-communication among different workers,  the following load-granularity constraint is applied.%the parameters in one neuron can not be allocated to different workers, yielding the following criterion.
\begin{granularity_constraint}[Neuron Allocation for ${\bf W}$-stage]\label{Criterion:WStage}
\emph{
In ${\bf W}$-stage, each neuron parametric vector (e.g., ${\bf w}_{g,i}$) defined in \eqref{Eq:WStep} should be allocated to one and only one worker.
}
\end{granularity_constraint}

\item ${\bf Z}$-stage: The auxiliary matrix ${\bf Z}$ is divided into $K$ blocks, with each being updated by one worker. To avoid inter-communication among workers, another load-granularity constraint is applied.
\begin{granularity_constraint}[Per-Sample Auxiliary Matrix Allocation for ${\bf Z}$-stage]\label{Criterion:ZStage}
\emph{
In ${\bf Z}$-stage, each per-sample auxiliary matrix (e.g., ${\bf Z}_m$) defined in  \eqref{Eq:ZStep} should be allocated to one and only one worker. 
}
\end{granularity_constraint}
\end{itemize}

\begin{example}\label{Rmk:SmallSizeSubproblem}
\emph{
Since the number of neurons in a DNN model and the data samples used for training are large, the sizes of each neuron problem and each per-sample auxiliary matrix problem are relatively small, compared with the whole learning tasks, making the model partitioning meaningful. As an example, our experiments involve the DNN model ``Lenet-5" proposed in \cite{lecun1998gradient} trained on the MNIST dataset. A mini batch of $50$ samples is used in each training iteration. In ``Lenet-5",  there are 3 convolutional layers, including $142$ feature maps in total. The first two convolutional layers are followed by a pooling layer and the last is followed by a fully connected layer with $84$ neurons. In ${\bf W}$-stage, the number of independent subproblems is $I = 142+84 = 226$. The size of each neuron problem is about $\dfrac{1}{I} = \dfrac{1}{226}$ of the whole problem. In ${\bf Z}$-stage, the size of each per-sample auxiliary matrix problem is $\dfrac{1}{50}$ of the whole problem.
}
\end{example}

Each stage (${\bf W}$-stage or ${\bf Z}$-stage) comprises three phases, push, computation, and pull, which are similar to those in the case of decomposable models. The main difference lies in the additional Granularity Constraint \ref{Criterion:WStage} or  \ref{Criterion:ZStage}. Each round comprises two stages and the rounds are repeated until the DNN model converges.

\subsection{Latency and Energy Consumption Models}

Consider an arbitrary communication round and an arbitrary worker, say worker $k$. The latency and energy consumption models of each phase are described as below.

\subsubsection{Push Phase} The push latency is the time for the server to broadcast the whole model-parameter vector to all workers. It is a constant identical for all workers. Besides, as the transmit power and bandwidth are very large during broadcasting, the push latency can be ignored. In this step, the energy consumption by all workers is to receive the model-parameter vector from the server and is included in the circuit energy consumption, denoted as $\xi$.

\subsubsection{Computation Phase}  The computation latency of worker $k$ depends on the size of the allocated parametric block $L_k$ and its computation speed $f_k$:
\begin{equation}\label{Eq:CompL}
T^{\rm cmp}_k = \dfrac{ L_k }{ f_k },\quad 1\leq k \leq K,
\end{equation}
where $f_k$ is measured by the number of parameters processed per second.

According to \cite{you2016energy}, the computation power of worker $k$ is 
%\begin{equation}
$P^{\rm cmp}_k = g_k f_k^3$, 
%\end{equation}
where $g_k$ is the computation power factor. Then, the computation energy of worker $k$ is 
\begin{equation}\label{Eq:CompE}
E^{\rm cmp}_k = P^{\rm cmp}_k \times T^{\rm cmp}_k= g_k f_k^2 L_k, \quad 1\leq k \leq K. 
\end{equation}

\subsubsection{Pull Phase} The pull phase consists of two parts. One is uploading gradient blocks from workers to the server. The other is the server updating the global model using the gradients sent by the workers. For the latter part, there is no energy consumption at the workers. Its latency, denoted as $T_{\rm s}$, is a constant and is same for all workers. In the sequel, we ignore the model update latency, $T_{\rm s}$, as it is small and has no impact on the solution of latency minimization.

For uploading, worker $k$ transmits over a set of assigned subcarriers. We denote $T^{\rm com}_{k,n}$ as the uploading latency of worker $k$ on subcarrier $n$. If subcarrier $n$ is not allocated to $k$, i.e., $C_{k,n}=0$, $T^{\rm com}_{k,n} = 0$. Otherwise, 
\begin{equation}\label{Eq:UploadLatencySubcarrier}
T^{\rm com}_{k,n}  = \dfrac{ L_{k,n} \tau }{ R_{k,n} },\quad \forall C_{k,n} =1,
\end{equation}
where $L_{k,n}$ is the number of parameters uploaded by worker $k$ on subcarrier $n$, $\tau$ is the number of bits per gradient element, and $R_{k,n}$ is the channel capacity of worker $k$ on subcarrier $n$. The channel capacity is given by
%\begin{equation}\label{Eq:SpectrumEfficiency}
$\left\{R_{k,n} = B \log_2\left( 1 +  P^{\rm com}_{k,n} h_{k,n} / \sigma^2  \right),\;\forall (k,n)\right\}$,
%\end{equation}
where $B$ is the subcarrier bandwidth, $\sigma^2$ is the power of additive white Gaussian noise, $P^{\rm com}_{k,n}$ is the transmit power, and $h_{k,n}$ is the channel gain of worker $k$ on subcarrier $n$, respectively. It follows that
\begin{equation}\label{Eq:Power}
P^{\rm com}_{k,n} = \dfrac{ \left( 2 ^{R_{k,n}/B} -1 \right) \sigma^2 }{ h_{k,n} },\quad \forall (k,n).
\end{equation}
Then, the overall uploading latency of worker $k$ is decided by the slowest subcarrier:
\begin{equation}\label{Eq:UploadL}
T^{\rm com}_k  = \max\limits_n\;\;T^{\rm com}_{k,n},\quad 1\leq k \leq K.
\end{equation}

The uploading energy consumption of worker $k$ is modeled as follows. Let $E^{\rm com}_{k,n}$ denote the transmit energy consumption of worker $k$ on subcarrier $n$. If subcarrier $n$ is not allocated, i.e., $C_{k,n}=0$, $E^{\rm com}_{k,n} =0$. Otherwise, 
\begin{equation}
E^{\rm com}_{k,n}  = C_{k,n}  P^{\rm com}_{k,n}  T^{\rm com}_{k,n} ,\quad \forall (k,n).%C_{k,n} =1.
\end{equation}
By substituting the transmit power density $P^{\rm com}_{k,n}$ in \eqref{Eq:Power} and the uploading latency $ T^{\rm com}_{k,n}$ in \eqref{Eq:UploadLatencySubcarrier}, $E^{\rm com}_{k,n}$ can be further derived as
\begin{equation}\label{Eq:UploadEnergySubcarrier}
E^{\rm com}_{k,n} = \dfrac{ C_{k,n} \left( 2 ^{R_{k,n}/B} -1 \right) \sigma^2   L_{k,n} \tau  }{ h_{k,n} R_{k,n}  },\quad \forall (k,n).
\end{equation}
The total uploading energy consumption of worker $k$ is the sum of uploading energy consumption over all subcarriers:
%\begin{equation}\label{Eq:EnergyCom}
$\left\{E^{\rm com}_{k} = \sum\nolimits_{n=1}^N E^{\rm com}_{k,n},\; 1\leq k \leq K\right\}$.
%\end{equation}
By substituting $E^{\rm com}_{k,n} $ in \eqref{Eq:UploadEnergySubcarrier}, 
\begin{equation}\label{Eq:UploadE}
E^{\rm com}_{k}= \sum\limits_{n=1}^N \dfrac{C_{k,n}  \left( 2 ^{R_{k,n}/B} -1 \right) \sigma^2   L_{k,n} \tau  }{ h_{k,n} R_{k,n}  }, \quad 1\leq k \leq K.
\end{equation}

Next, the total latency and energy consumption of worker $k$ are defined as follows. The latency of worker  $k$ is the sum latencies of the two phases:
\begin{equation}\label{Eq:WorkerL}
T_k =  T^{\rm cmp}_k+ T^{\rm com}_k,\quad 1\leq k \leq K,
\end{equation}
where $ T^{\rm cmp}_k$ is the computation latency defined in \eqref{Eq:CompL}, $T^{\rm com}_k$ is the uploading latency defined in \eqref{Eq:UploadL}. The energy consumption of worker $k $ is given by:
\begin{equation}\label{Eq:WorkerE}
E_k = E^{\rm cmp}_k+ E^{\rm com}_k + \xi,\quad 1\leq k \leq K,
\end{equation}
where $\xi$ is the constant circuit energy consumption when there is no computation and transmission, $E^{\rm cmp}_k$ defined in \eqref{Eq:CompE} and $E^{\rm com}_k$ defined in \eqref{Eq:UploadE} are the computation and uploading energy consumption of worker $k$, respectively.

\section{Problem Formulation}

We aim at minimizing the overall learning latency of the PARTEL system, which depends on two factors: the per-round latency and the number of rounds for model convergence. The overall learning latency is defined as the total latency of all rounds till model convergence. In \cite{wen2020joint} for narrowband channels, it is proved that the overall learning latency minimization is equivalent to separately minimizing the per-round latency. The result can also apply to the current case of   broadband channels, as stated below.
\begin{lemma}[Equivalent Per-Round Latency Minimization \cite{wen2020joint}]\label{Lma:PerRoundL}
\emph{
The overall learning latency minimization is equivalent to separately minimizing the latencies for all rounds.
}
\end{lemma}
Lemma \ref{Lma:PerRoundL} holds because the distributed learning algorithms implemented using PARTEL are equivalent to the corresponding centralized ones in terms of convergence rate as measured by the required number of communication rounds. Specifically, for distributed learning, the values of updates (e.g., gradients and proximal gradients) calculated in each round and the number of rounds required for model convergence are independent of SUPPORT.

Using the result in Lemma \ref{Lma:PerRoundL}, we formulate the equivalent per-round latency-minimization problem. For an arbitrary round, we aim to minimize its latency, denoted as $T$, 
%The objective, say the per-round latency, is decided by the slowest worker in the system: 
%\begin{equation}\label{Eq:SystemL}
%t  = \max\limits_k\;\; T_k,
%\end{equation}
%where $T_k$ defined in \eqref{Eq:WorkerL} is the latency of worker $k$.
%We aim at minimizing the per-round latency 
under the constraints on subcarrier assignment, latency requirement, parameter  allocation, and power control, described as follows.

\subsubsection{Subcarrier Assignment Constraints} 
Each subcarrier can be allocated to one worker:
\begin{equation}(\text{C1: Subcarrier Assignment Constraint})\quad
\left\{
\begin{aligned}
&C_{k,n} \in \{0,1\}, \quad \forall (k,n),\\
&\sum\limits_{k=1}^K C_{k,n} = 1,\quad 1\leq n\leq N,
\end{aligned}
\right.
\end{equation}
where $C_{k,n} = 1$ represents that the subcarrier $n$ is allocated to worker $k$. 

\subsubsection{Per-Round Latency Constraints} 
As all parametric blocks should be updated in one round, all workers' latencies, say $\{T_k\}$, should not exceed the overall one-round latency $T$:
\begin{equation}\quad\label{Eq:LatencyConstraint1}
T_k \leq T, \quad 1\leq k \leq K.
\end{equation}
As mentioned, $T$ is the latency for an arbitrary round and can be different over different rounds.
By substituting $T_k$ in \eqref{Eq:WorkerL}, the constraints in \eqref{Eq:LatencyConstraint1} can be derived as
\begin{equation}
 T^{\rm cmp}_k +  T^{\rm com}_k \leq T, \quad 1\leq k \leq K,
\end{equation}
which, by substituting the uploading latency $T^{\rm com}_k$ in \eqref{Eq:UploadL}, are equivalent to
\begin{equation}(\text{C2: Per-Round Latency Constraint})\quad
 T^{\rm cmp}_k +  T^{\rm com}_{k,n} \leq T, \quad \forall C_{k,n}=1,
\end{equation}
where $T^{\rm cmp}_k$ defined in \eqref{Eq:CompL} is the computation latency of worker $k$ and $T^{\rm com}_{k,n}$ defined in \eqref{Eq:UploadLatencySubcarrier} is the uploading latency of worker $k$  on subcarrier $n$. 

\subsubsection{Parameter Constraints} 
The parameter constraints are two tiers. On the one hand, the total updatable number of parameters by all workers should be no smaller than the size of the model:
\begin{equation}(\text{C3: Inter-Worker Parameter Constraint})\quad
\sum\limits_{k=1}^K L_k \geq L,
\end{equation}
where $L_k$ is the size of the parametric block allocated to worker $k$ and $L$ is the size of the model-parameter vector (or matrix). On the other hand, for each worker, the total uploaded number of parameters on all subcarriers should be no smaller than its allocated parametric-block size:
\begin{equation}\label{Eq:WParameterConstraint}(\text{C4: Intra-Worker Parameter Constraint})\quad
\sum\limits_{n=1}^N C_{k,n} L_{k,n} \geq L_k,\quad 1\leq k \leq K,
\end{equation}
where $L_{k,n}$ is the number of parameters uploaded by worker $k$ on subcarrier $n$. In the sequel,  $\{L_k\}$ and $\{L_{k,n}\}$ are relaxed to be continuous for simplicity. In practice, the solved $\{L_k^*\}$ and $\{L_{k,n}^*\}$ will be rounded for implementation and the loss caused by the rounding operation can be ignored, since the values of $\{L_k\}$ and $\{L_{k,n}\}$ are typically large. 

For the case of DNN models, Granularity Constraints \ref{Criterion:WStage} and \ref{Criterion:ZStage} can be written mathematically as follows.
\begin{equation}\label{Eq:WParameterConstraint}({\rm C_{dnn}}\text{: Additional Parameter Constraint for DNN Models})\quad
\dfrac{L_k}{L_{\rm sub}} \in \mathbb{N}^+,\quad 1\leq k\leq K,
\end{equation}
where $\mathbb{N}^+$ is the set of positive integers and $L_{\rm sub}$ is the size of the subproblems, i.e., neurons or per-sample auxiliary matrices. For ${\bf W}$-stage, the size of all neurons, say $L_{\rm sub}$, are assumed the same for simplicity, which has little impact on the solution, since the size of each neuron is much smaller than that of the whole problem, as mentioned in Example \ref{Rmk:SmallSizeSubproblem}. For ${\bf Z}$-stage, the size of each per-sample auxiliary matrix is the total number of neurons, say $L_{\rm sub}=\sum_{g=1}^G I_g$ with $I_g$ being the number of neurons in layer $g$.

%are currently not considered. But it will be used in Section \ref{Section:DesignDNN} to derive a practical communication-efficient scheme for DNN models.

\subsubsection{Power Constraints} 
The power consumption of each worker is constrained as
\begin{equation}(\text{C5: Power Constraint})\quad
\dfrac{ E_k  } { T_k  } \leq P_k,\quad 1\leq k \leq K,
\end{equation}
where $E_k $ defined in \eqref{Eq:WorkerE}, $T_k  $ defined in \eqref{Eq:WorkerL}, and $P_k$ are the energy consumption, latency, and maximal permitted power of worker $k$, respectively. 

\subsubsection{Latency-Minimization Problem}
Under these constraints, the per-round latency-minimization problem by joint SUPPORT can be formulated as
\begin{equation}\text{({P1})}\quad
\begin{aligned}
\mathop{\min }\limits_{\{C_{k,n}\}, \{L_k\}, \{L_{k,n}\}, \{R_{k,n}\},T }\;\;  &T,\\
{\text{s.t.}}\;\; \text{(C1)} \sim \text{(C5)},\; & \& \; ({\rm C_{dnn}})\text{ for a DNN Model}.%&C_{k,n} \in \{0,1\}, \quad \forall (k,n),\\
%&\sum\limits_{k=1}^K C_{k,n} = 1,\quad 1\leq n\leq N,\\
%&T^{\rm cmp}_k +  T^{\rm com}_{k,n} \leq T, \quad \forall C_{k,n}=1,\\
%&\sum\limits_{k=1}^K L_k= L,\\
%&\sum\limits_{n=1}^N C_{k,n} L_{k,n} = L_k,\quad 1\leq k \leq K,\\
%&\dfrac{ E_k  } { T_k  } \leq P_k,\quad 1\leq k \leq K.
\end{aligned}
\end{equation}

\section{Joint SUPPORT for Decomposable Models}\label{Section:DesignDecomposable}

In this section, joint SUPPORT is designed by developing a tractable approach for solving Problem (P1).

\subsection{Equivalent Latency Requirement}

First, the following necessary condition for the equivalent latency requirement can be derived to simplify Problem (P1). Note that in \cite{wen2020joint}, similar equivalent latency property can be derived in the PARTEL design for  frequency non-selective channels. However, for OFDM systems considered in this paper, the binary subcarrier allocation among workers and the corresponding inter-subcarrier power and parameter allocation for each device make the problem much more complicated.

\begin{lemma}[Equivalent Latency for All Workers]\label{Lma:EquivalentL}
\emph{ 
To achieve the optimal solution of (P1), the following latency condition should be satisfied:
\begin{equation}\label{Eq:LatencyNecessaryCondition}
T^{\rm cmp}_k +  T^{\rm com}_{k,n} = T, \quad \forall C_{k,n}=1,
\end{equation}
where  $T^{\rm cmp}_k$ defined in \eqref{Eq:CompL} is the computation latency of worker $k$, $T^{\rm com}_{k,n} $ defined in \eqref{Eq:UploadLatencySubcarrier} is the uploading latency of worker $k$  on subcarrier $n$.
}
\end{lemma}
\proof See Appendix \ref{Apdx:LmaEquivalentL}.

The result in Lemma \ref{Lma:EquivalentL} yields the following insights. First, it requires all workers the same latency with the overall latency $T$. Second, for each worker, the uploading latency on all allocated subcarriers should be equal.

\begin{remark}[Computation Latency vs. Communication Latency] \emph{
By substituting the computation latency $T^{\rm cmp}_k $ in \eqref{Eq:CompL} and the communication latency $T^{\rm com}_{k,n}$ in \eqref{Eq:UploadLatencySubcarrier}  into the equivalent latency property in \eqref{Eq:LatencyNecessaryCondition}, it can be derived as
\begin{equation}\label{Eq:Remark1}
 \dfrac{ L_k }{ f_k } + \dfrac{ L_{k,n} \tau }{ R_{k,n} } = T,\quad \forall C_{k,n}=1,
\end{equation}
with the constraints $\sum\nolimits_{n=1}^N C_{k,n}L_{k,n}=L_k$. From \eqref{Eq:Remark1}, the load, say $L_k$, has the same effect on computation and communication latency. On the other hand, when the computation frequency $f_k$ is small compared to the number of subcarriers and the data rates, the computation latency dominates or vice versa. 
}
\end{remark}

By substituting the computation latency $T^{\rm cmp}_k$ defined in \eqref{Eq:CompL}  and the uploading latency $T^{\rm com}_{k,n}$ defined in \eqref{Eq:UploadLatencySubcarrier} into the necessary condition in Lemma \ref{Lma:EquivalentL}, we can derive the number of parameters uploaded by worker $k$ on subcarrier $n$, say $L_{k,n}$, as
\begin{equation}\label{Eq:SubcarrierParameter}
L_{k,n} = \dfrac{ C_{k,n} R_{k,n} }{  \tau }\left( T -  \dfrac{ L_k }{f_k} \right),\quad \forall (k,n),
%L_{k,n}= \left\{
%\begin{aligned}
%&\dfrac{ R_{k,n} }{  \tau }\left( T -  \dfrac{ L_k }{f_k} \right),\quad &\text{if } C_{k,n} =1,\\
%&0, \quad &\text{otherwise},
%\end{aligned}
%\right.
\end{equation}
where $T$ is the per-round latency, $C_{k,n}\in\{0,1\}$ is the subcarrier-allocation indicator, $L_k$ is the parametric-block size allocated to worker  $k$, $R_{k,n}$ is the channel capacity of $k$ on subcarrier $n$. By substituting $L_{k,n}$ defined in \eqref{Eq:SubcarrierParameter} and the necessary condition in Lemma \ref{Lma:EquivalentL}, Problem (P1) can be simplified as:
\begin{equation}\text{(P2)}\quad
\begin{aligned}
\mathop{\min }\limits_{ \substack{ \{C_{k,n}\}, \{L_k\}, \\ \{R_{k,n}\},T }}&\;\;  T,\\
{\text{s.t.}}\;\; &\text{(C1), (C3)},\\
%&C_{k,n} \in \{0,1\}, \quad \forall (k,n),\\
%&\sum\limits_{k=1}^K C_{k,n} = 1,\quad 1\leq n\leq N,\\
%&\sum\limits_{k=1}^K L_k\geq L,\\
&\sum\limits_{n=1}^N \dfrac{C_{k,n} R_{k,n}  }{  \tau }\left( T -  \dfrac{ L_k }{f_k} \right) \geq L_k,\quad 1\leq k \leq K,\\
& E_k   \leq P_k T,\quad 1\leq k \leq K,
\end{aligned}
\end{equation}
where $E_k  $ defined in \eqref{Eq:WorkerE} is the energy consumption of worker $k$. By substituting $L_{k,n}$ in \eqref{Eq:SubcarrierParameter}, $E_k  $  can be expressed as
\begin{equation}\label{Eq:WorkerE1}
E_k  = g_kf_k^2L_k + \sum\limits_{n=1}^N  \dfrac{ C_{k,n} \left( 2^{ R_{k,n}/B }-1 \right)  \sigma^2 }{ h_{k,n} } \left( T -   \dfrac{ L_k }{f_k} \right) +  \xi .
\end{equation}
%To make it more tractable, the equalities of the two parameter constraints are changed to inequalities in Problem (P2), which has no impact on the optimality. %Problem (P2) is a mixed integer non-convex problem and is NP-hard. To solve it, we follow the typical way (see, e.g., \cite{wong1999multiuser}), which first relaxes the subcarrier-allocation indicators to $\left\{C_{k,n}\in [0,1],\;\forall (k,n)\right\}$ to solve the approximated problem. Then, based on approximated solution, the near-optimal solution of Problem (P2) is found. 

\subsection{Equivalent Convex Problem}

Problem (P2) is a mixed integer non-convex problem and is hence NP-hard \cite{burer2012non}. In the sequel, two steps are used to tackle it. First, following the standard approach to tackle integer programming (see e.g., \cite{wong1999multiuser}),  linear programming relaxation is used to relax the subcarrier-allocation indicators in Problem (P2) to be continuous, i.e., $\left\{C_{k,n}\in [0,1],\;\forall (k,n)\right\}$. Then, following the method in \cite{wen2020joint}, the relaxed problem can be equivalently converted to the problem of updatable model size maximization.  However, it remains non-convex and difficult to tackle due to the intra-worker parameter constraint and the power constraint. In the sequel, the problem of updatable model size maximization is derived and solved.

Given the one-round latency $T$ for an arbitrary round, let $\hat{L}^*(T)$ denote the maximum size of a model that can be updated within the round.  Then $\hat{L}^*(T)$ solves the following problem of model size maximization:
\begin{equation}(\text{P3})\quad
\begin{aligned}
\hat{L}^*(T) = \mathop{\max }\limits_{\{C_{k,n}\}, \{L_k\}, \{R_{k,n}\}}\;\;  &\sum\limits_{k=1}^K L_k,\\
{\text{s.t.}}\;\; &0\leq C_{k,n} \leq 1, \quad \forall (k,n),\\
&\sum\limits_{k=1}^K C_{k,n} = 1,\quad 1\leq n\leq N,\\
&\sum\limits_{n=1}^N \dfrac{C_{k,n} R_{k,n}  }{  \tau }\left( T -   \dfrac{ L_k }{f_k} \right) \geq L_k,\quad 1\leq k \leq K,\\
& %G f_k^2 L_k + \sum\limits_{n=1}^N \dfrac{C_{k,n}  \left( 2 ^{R_{k,n}} -1 \right) \sigma^2   L_{k,n} \tau  }{ h_{k,n} R_{k,n}  }  \leq P_k t,\quad 1\leq k \leq K.
E_k   \leq P_k T,\quad 1\leq k \leq K,
\end{aligned}
\end{equation}
where $C_{k,n}$ is the subcarrier-allocation indicator, $L_k$ is the parametric-block size allocated to worker $k$, $R_{k,n}$ is the channel capacity of worker  $k$ on subcarrier $n$, $T$ is the one-round latency, $E_k $ defined in \eqref{Eq:WorkerE1} is the energy consumption of worker  $k$. Note that solving Problem (P2) via utilizing the problem of model size maximization in Problem (P3) follows the method in \cite{wen2020joint}. However, new challenges arise from the subcarrier allocation among workers and the inter-subcarrier power and parameter allocation for each worker, leading to the non-convexity and a much larger size of Problem (P3). 

\begin{lemma}[Relation of Maximal Model Size and Latency]\label{Lma:RelationModelSizeLatency}
\emph{
$\hat{L}^*(T)$ defined in Problem (P3) is a monotonously increasing function of $T$.
}
\end{lemma}
\proof See Appendix \ref{Apdx:LmaRelationModelSizeLatency}.

It follows from the result in Lemma \ref{Lma:RelationModelSizeLatency} that the solution of Problem (P2) is the minimal latency, say $T^*$, which makes  the updatable model size $\hat{L}^*(T^*)$ no less than the target size  $L$. This suggests a method to solve Problem (P2) by searching $T^*$ using the criterion $\hat{L}^*(T)\geq L$, which will be elaborated in the later subsection. 

To get the maximum updatable model size $\hat{L}^*(T)$ requires solving Problem (P3). To this end, the following variables  are used to transform Problem (P3) into a \emph{convex} problem.
\begin{equation}\label{Eq:VariableTransform}
\left\{
\begin{aligned}
& \phi_k = \left( T -  \dfrac{ L_k }{f_k} \right)^{-1},\\
& \tilde{R}_{k,n} = C_{k,n} R_{k,n},
\end{aligned}
\right.
\end{equation}
By substituting the variables in \eqref{Eq:VariableTransform} and $E_k$ defined in \eqref{Eq:WorkerE1}, Problem (P3) can be written as 
\begin{equation*}(\text{P4})\;
\begin{aligned}
\hat{L}^*(T) =  \max\limits_{\substack { \{C_{k,n}\}, \{\phi_k\} ,\\ \{\tilde{R}_{k,n}\},} }& \sum\limits_{k=1}^K  f_k \left( T - \dfrac{ 1 }{ \phi_k } \right),\\
\text{s.t.}\;\; & 0\leq C_{k,n}\leq 1,\quad \forall(k,n),\\
& \sum\limits_{k=1}^K C_{k,n} = 1,\quad 1\leq n\leq N,\\
& \sum\limits_{n=1}^N \dfrac{ \tilde{R}_{k,n} }{  \tau } \geq  f_k \left( T \phi_k - 1 \right),\quad 1\leq k\leq K,\\
& \sum\limits_{n=1}^N  \dfrac{C_{k,n}  \sigma^2 \big( 2^{ \frac{\tilde{R}_{k,n} }{BC_{k,n} } }-1 \big) }{ h_{k,n} }   +  g_kf_k^3 \left( \phi_k T- 1 \right) \leq ( P_kT - \xi) \phi_k,\; 1\leq k\leq K.
%\sum\limits_{n=1}^N  \dfrac{C_{k,n}  \sigma^2 \big( 2^{ \frac{\tilde{R}_{k,n} }{C_{k,n} } }-1 \big) }{ h_{k,n} (t - T_{\rm ph} - T_{\rm s}) }   +  g_kf_k^3 \left[ \phi_k - \dfrac{1}{t - T_{\rm ph} - T_{\rm s}} \right] \leq P_k\phi_k,\; 1\leq k\leq K.
\end{aligned}
\end{equation*}

\begin{lemma}\label{Lma:Convexity}
\emph{
Problem (P4) is a convex problem.
}
\end{lemma}
\proof See Appendix \ref{Apdx:LmaConvexity}.

\subsection{Properties of Optimal Policies}

Based on the results in the previous subsection, the optimal policies of Problem (P2) with relaxed subcarrier-allocation indicators are proposed, as described in the following.

As (P4) is convex, the primal-dual method can be used to get the optimal solution:
\begin{equation}
\max\limits_{\substack{ \{\mu_n\},\{ \lambda_k\},\\ \{\nu_k\} }}\; \min\limits_{\substack{\{C_{k,n}\},\{\tilde{R}_{k,n}\},\\\{\phi_k\}}} \;\; \mathcal{L_{\text{P4}}},
\end{equation}
where $\mathcal{L_{\text{P4}}}$ is the Lagrange function of Problem (P4), given as 
\begin{equation}\label{Eq:LagrangeFunctionP4}
\begin{aligned}
\mathcal{L_{\text{P4}}} =& - \sum\limits_{k=1}^K f_k \left( T - \dfrac{ 1 }{ \phi_k } \right)+ \sum\limits_{n=1}^N \mu_n \left(  1 - \sum\limits_{k=1}^K C_{k,n}  \right) + \sum\limits_{k=1}^K \lambda_k \left[ f_k \left( T\phi_k - 1 \right) - \sum\limits_{n=1}^N \dfrac{  \tilde{R}_{k,n} }{  \tau } \right] \\
&+  \sum\limits_{k=1}^K\nu_k \left[ \sum\limits_{n=1}^N C_{k,n} \left( 2^{ \frac{\tilde{R}_{k,n}}{BC_{k,n} } }-1 \right) \times \dfrac{  \sigma^2 }{ h_{k,n} }  + g_kf_k^3 \left( T\phi_k - 1 \right) - (P_kT-\xi) \phi_k \right],
\end{aligned}
\end{equation}
and $\{\mu_n\}$, $\{\lambda_k\geq 0\}$, and $\{\nu_k\geq 0\}$ are Lagrangian multipliers.  

Next, the necessary conditions for achieving the optimal solution of the inner loop are used  to derive the optimal policies. The inner loop problem is given by
\begin{equation}\label{Eq:InnerLoop}
\min\limits_{ \{C_{k,n}\},\{\tilde{R}_{k,n}\},\{\phi_k\}} \;\; \mathcal{L_{\text{P4}}},\quad \text{given } \{\mu_n\}, \{\lambda_k\}, \{\nu_k\}.
\end{equation}
%the following conditions are necessary to achieve the optimal solution. 

%\subsubsection{Optimal Power Allocation} 
The first necessary condition  is 
\begin{equation}
\dfrac{ \partial \mathcal{L_{\text{P4}}} }{ \partial \tilde{R}_{k,n} } = -\dfrac{ \lambda_k }{ \tau } + \nu_k 2^{ \frac{\tilde{R}_{k,n}}{BC_{k,n} } }\ln2 \times \dfrac{  \sigma^2 }{ Bh_{k,n} }=0,\quad \forall C_{k,n}\neq0,
\end{equation}
which gives the following optimal scheme for calculating the channel capacity: 
\begin{equation}\label{Eq:OptimalChannelCapacity}
R_{k,n}^* = \left\{
\begin{aligned}
&\dfrac{ \tilde{R}_{k,n}^* }{ C_{k,n}^* }  = B\log_2 \left( \dfrac{ \lambda_k B }{ \nu_k\tau \ln2 } \right)+ B\log_2 \left( \dfrac{ h_{k,n} }{  \sigma^2 } \right),\quad &\forall C_{k,n}\neq 0, \\
&0, \quad &\text{otherwise}.
\end{aligned}
\right.
\end{equation}
By substituting $R_{k,n}^*$ in \eqref{Eq:OptimalChannelCapacity} into the transmission power in \eqref{Eq:Power}, the optimal power-allocation scheme can be derived, as in the following lemma.

\begin{lemma}[Optimal Power Allocation]
\emph{
The optimal power-allocation scheme is
\begin{equation}\label{Eq:OptimalPowerAllocation}
{P^{\rm com}_{k,n}}^*= \left\{
\begin{aligned}
& \dfrac{ \lambda_k B }{ \nu_k\tau \ln2 } - \dfrac{\sigma^2 }{ h_{k,n} },\quad &\forall C_{k,n}\neq 0, \\
& 0,\quad &\text{otherwise}.
\end{aligned}
\right.
\end{equation}
}
\end{lemma}

The water-filling like result in \eqref{Eq:OptimalPowerAllocation} shows that for each worker, more power should be allocated on the subcarrier with high channel gain, say $h_{k,n}$.

%\subsubsection{Optimal Parameter Allocation}
The second necessary condition to achieve the optimum of the inner loop problem in \eqref{Eq:InnerLoop} is 
\begin{equation}\label{Eq:SecondCondition}
\dfrac{ \partial \mathcal{L_{\text{P4}}} }{ \partial \phi_k } = -\dfrac{ f_k }{ \phi_k^2 } + \lambda_k f_k T + \nu_kg_kf_k^3 T - \nu_k(P_kT-\xi) = 0. 
\end{equation}
By substituting the variable transformations in \eqref{Eq:VariableTransform} into \eqref{Eq:SecondCondition}, we can achieve the optimal inter-worker parameter allocation scheme, as follows.
\begin{lemma}[Optimal Parameter Allocation among Workers]\label{Lma:PAW}
\emph{
The optimal inter-worker parameter-allocation scheme is
\begin{equation}\label{Eq:OptimalWorkerParameter}
L_k^* =\left[ T- \sqrt{  \lambda_k  T + \nu_kg_kf_k^2T - \nu_k (P_k T -\xi)/f_k} \right] f_k,\quad 1\leq k \leq K.
\end{equation}
}
\end{lemma}
In \eqref{Eq:OptimalWorkerParameter}, the optimal parametric-block size allocated to worker $k$, say $L_k^*$, is a concave function of the computation speed $f_k$ and a monotone decreasing function of the computation power factor $g_k$. On one hand, large $f_k$ can reduce the computation latency. On the other hand, the computation energy increases as a square function of $f_k$. The optimal load in \eqref{Eq:OptimalWorkerParameter} balances the two aspects.

Substituting the parameter-allocation scheme in \eqref{Eq:OptimalWorkerParameter} and the channel capacity in \eqref{Eq:OptimalChannelCapacity} into the intra-worker parameter-allocation scheme $\{L_{k,n}\}$ in \eqref{Eq:SubcarrierParameter}, gives the following lemma.

\begin{lemma}[Optimal Parameter Allocation Among Subcarriers]
\emph{
The optimal intra-worker parameter allocation scheme is given by
\begin{equation}\label{Eq:ParameterAllocationSubcarrier}
L_{k,n}^*=\left\{
\begin{aligned}
&\dfrac{\sqrt{  \lambda_k T + \nu_kg_kf_k^2 T - \nu_k (P_k T -\xi)/f_k  }}{\tau }\times  B\log_2 \left( \dfrac{ \lambda_k B h_{k,n} }{ \nu_k\tau \sigma^2   \ln2 }  \right),\;&\text{if } C_{k,n}\neq 0,\\
& 0, \; &\text{otherwise}.
\end{aligned}
\right.
\end{equation}
%where  $\{ \lambda_k\geq 0 \}$ and $\{ \nu_k \geq 0 \}$ are multipliers. 
}
\end{lemma}
%\begin{remark}[Intra-Worker Parameter Allocation versus Channel Gain]
%\emph{
From \eqref{Eq:ParameterAllocationSubcarrier}, more parameters should be assigned to the channel with high gain.
%}
%\end{remark}

%\subsubsection{Optimal Subcarrier Allocation}
The third necessary condition to achieve the optimum of the inner loop problem in \eqref{Eq:InnerLoop} is 
\begin{equation}\label{Eq:ThirdCondition} 
\dfrac{ \partial \mathcal{L_{\text{P4}}} }{ \partial C_{k,n} }  = -\mu_n +   I_{k,n} = 0, \quad \forall (k,n),
\end{equation}
where $I_{k,n}$ is the indicator function given by
\begin{equation}\label{Eq:IndicatorFunction}
I_{k,n} = \dfrac{ \nu_k \sigma^2 }{ h_{k,n} } \left[ \left( 2^{R_{k,n} ^*/B} -1 \right)  -   \dfrac{R_{k,n} ^*2^{ R_{k,n} ^*/B } \ln2}{B}   \right],\quad \forall (k,n).
\end{equation}
Note that $I_{k,n}$ is determined when $R_{k,n} ^*$ is known. Let $\mu_n = \min\limits_k I_{k,n}$. If $I_{k,n}> \mu_n$, $C_{k,n}=0$, as the condition in \eqref{Eq:ThirdCondition} can not be satisfied. If $I_{k,n}= \mu_n$ for a unique worker, say $k$, then $C_{k,n}=1$. If $I_{k,n}= \mu_n$ for multiple workers, then $C_{k,n}\in (0,1)$ for these workers. And in the last case, it is easy to show that the values of the non-zero $\{C_{k,n}\}$ won't influence the value of the Lagrange function $\mathcal{L_{\text{P4}}}$ defined in \eqref{Eq:LagrangeFunctionP4}, as long as the subcarrier assignment constraint, say $\left\{ \sum\nolimits_{n=1}^N C_{k,n} = 1,\quad 1\leq n \leq N\right\}$, are satisfied. 

The optimal subcarrier allocation  is summarized in the following lemma.

\begin{lemma}[Optimal Subcarrier Allocation]\label{Lma:SA}
\emph{
The optimal subcarrier allocation is given as:
\begin{equation}\label{Eq:OptimalSubcarrierAllocation}
C_{k,n}^*  \left\{
\begin{aligned}
&=0,\qquad\;\text{if } I_{k,n}>\mu_n,\\
&\in(0,1),\;\;\text{if } I_{k,n}=\mu_n \text{ for multiple workers},\\
&=1,\qquad\;\text{if }  I_{k,n}=\mu_n \text{ for a unique worker}  \mathcal{W}_k,
\end{aligned}
\right.
\end{equation}
where $I_{k,n} $ is the indicator function defined in \eqref{Eq:IndicatorFunction}, $\mu_n = \min\limits_k\; I_{k,n}$, and $ R_{k,n}^*$ is the optimal channel capacity in \eqref{Eq:OptimalChannelCapacity}. %Besides, for the second case in \eqref{Eq:OptimalSubcarrierAllocation}, the values of the non-zero $\{C_{k,n}\}$ won't influence the optimality, as long as the subcarrier constraints, say $\left\{ \sum\nolimits_{n=1}^N C_{k,n} = 1,\quad 1\leq n \leq N\right\}$, are satisfied.
}
\end{lemma}
%\begin{remark}[Subcarrier Allocation versus Channel Gain]
%\emph{
In \eqref{Eq:OptimalSubcarrierAllocation}, a high channel gain leads to a small value of $I_{k,n}$ and thus a high possibility to make $C_{k,n}\neq 0$. That means the subcarrier with higher channel gain has larger possibility to be allocated to the worker.
%}
%\end{remark}
Note that in the optimal scheme in Lemma \ref{Lma:SA}, some subcarrier-allocation indicators may be fractions. The standard approach is to round those to be binary (see, e.g., \cite{wong1999multiuser}), which will be elaborated in the later subsection.

\subsection{Optimal Policy Computation}

In this subsection, the joint SUPPORT algorithm to solve the original Problem (P1) is proposed. 
First, we solve the convex Problem (P4) by the primal-dual method using the closed-form results in Lemmas \ref{Lma:PAW}-\ref{Lma:SA}. Some notation is described as follows. $\{\eta_{\lambda_k}\}$ and $\{\eta_{\nu_k}\}$ denote the step sizes of gradient descent. $\mathcal{L}_{\text{P4}}$ and $\mu$, $\{\lambda_k\geq 0\}$, and $\{\nu_k\geq 0\}$ are the Lagrange function and Lagrangian multipliers defined in \eqref{Eq:LagrangeFunctionP4}. With the notation, the application of the primal-dual method yields Algorithm \ref{Ag:Feasibility} for solving Problem (P4). 
\begin{remark}[Low Complexity of Updatable Model Size Maximization]
\emph{
The computation complexity of Algorithm \ref{Ag:Feasibility} is $\mathcal{O}(K^2N)$ with $K$ being the number of workers and $N$ being the number of subcarriers, as the closed-form results in Lemmas \ref{Lma:PAW} - \ref{Lma:SA} makes the updating of corresponding variables more efficient. As a comparison, directly solving the non-convex Problem (P3) has a computational complexity of at least $\mathcal{O}(K^3N^3)$ and is suboptimal.
}
\end{remark}

\begin{algorithm}[t]
\caption{Updatable Model Size Maximization}\label{Ag:Feasibility}
%\begin{algorithmic}[1]

1: {\bf Input:} channel gains $\{h_{k,n}\}$, computation speeds $\{f_k\}$, computation power factors, $\{g_k\}$, and the given one-round latency $T$.
%\begin{itemize}
%\item $\{h_{k,n}\}$, the channel gains,
%\item $T$, the given one-round latency.
%\end{itemize}
  
2: {\bf Initialize} $\{\lambda_k^{(0)}\}$, $\{\nu_k^{(0)}\}$, and $i=0$.

3: {\bf Loop}

4: \;\;\;\;Update the multipliers as 
\begin{equation*}
\left\{
\begin{aligned}
&\lambda_k^{(i+1)} = \max\left\{\lambda_k^{(i)} +\eta_{\lambda_k} \dfrac{ \partial \mathcal{L}_{\text{P4} }}{\partial \lambda_k}, \quad 0\right\}, \;1\leq k \leq K, \\
&\nu_k^{(i+1)} = \max\left\{\nu_k^{(i)} +\eta_{\nu_k} \dfrac{ \partial \mathcal{L}_{\text{P4} }}{\partial \nu_k}, \quad 0\right\}, \;1\leq k \leq K,
\end{aligned}
\right.
\end{equation*}

5: \;\;\;\;Solve $\{L_k^*\}$, $\{R_{k,n}^*\}$, and $\{C_{k,n}^*\}$ using \eqref{Eq:OptimalWorkerParameter}, \eqref{Eq:OptimalChannelCapacity}, and \eqref{Eq:OptimalSubcarrierAllocation}, respectively.

6: \;\;\;\;Get $\{\phi_k^*\}$ and $\{\tilde{R}_{k,n}^*\}$ with \eqref{Eq:VariableTransform}.

7: {\bf Until Convergence}

8: $\hat{L}^*(T) = \sum\nolimits_{k=1}^K L_k^*$.

9: {\bf Output:} $\hat{L}^*(T)$, $\{L_k^*\}$, $\{R_{k,n}^*\}$, and $\{C_{k,n}^*\}$.

\end{algorithm}

Then, as mentioned in the preceding subsection, Problem (P2) with relaxed subcarrier-allocation indicators can be solved by nesting a one-dimensional search over the latency $T$ and solving the convex Problem (P4). Based on the monotonicity of $\hat{L}^*(T)$ in Lemma \ref{Lma:RelationModelSizeLatency}, the search can be efficiently implemented by bisection method. While the solution of Problem (P4) is presented in Algorithm \ref{Ag:Feasibility}. Then the optimal policy to solve Problem (P2) with relaxed subcarrier-allocation indicators is presented in Algorithm \ref{Ag:JointOptimalDesign}, by nesting the bisection search and Algorithm \ref{Ag:Feasibility}.

\begin{algorithm}[t]
\caption{Joint SUPPORT}\label{Ag:JointOptimalDesign}
1: {\bf Input:} channel gains $\{h_{k,n}\}$, computation speeds $\{f_k\}$, and computation power factors, $\{g_k\}$.
%\begin{itemize}
%\item $\{h_{k,n}\}$, the channel gains.
%\end{itemize}

2: {\bf Select} $T = T_{\rm u}$ that makes $\hat{L}^*(T_{\rm u})$  defined in Problem ({P4}) larger than $L$.

3:  {\bf Select} $T = T_{\rm l}$ that makes $\hat{L}^*(T_{\rm l})<L$.

4: {\bf While} $T_{\rm u }\not= T_{\rm l}$

5: \;\;\;\;Let $T_{\rm m} = (T_{\rm u}+T_{\rm l})/2$. 

6: \;\;\;\;Input $\{h_{k,n}\}$, $\{f_k\}$, $\{g_k\}$ and $T=T_ {\rm m}$  into Algorithm \ref{Ag:Feasibility} to solve ({P4}).

7: \;\;\;\;Obtain  $\hat{L}^*(T_{\rm m})$, $\{L_k^*\}$, $\{R_{k,n}^*\}$, and $\{C_{k,n}^*\}$.

8: \;\;\;\;{\bf If} $\hat{L}^*(T_{\rm m})\geq L$

9: \;\;\;\;\;\;\;\;$T_{\rm u} =  T_{\rm m}$.

10:\;\;\;\;{\bf Else}

11:\;\;\;\;\;\;\;\;$T_{\rm l} = T_{\rm m}$.

12:\;\;\;\;{\bf End if}

13:{\bf End while}

14:$T^* = T_{\rm m}$.

15:{\bf Output:} $T^*$, $\{L_k^*\}$, $\{R_{k,n}^*\}$, and $\{C_{k,n}^*\}$.
\end{algorithm}

Finally, based on Algorithm \ref{Ag:JointOptimalDesign}, the joint scheme of SUPPORT without relaxation is proposed to solve the original Problem (P1). Note that not all subcarrier-allocation indicators solved by Algorithm \ref{Ag:JointOptimalDesign} are integers, i.e.,  $C_{k,n}^* \in (0,1)$ for some $(k,n)$. For these subcarriers, a practical subcarrier-allocation scheme following \cite{wong1999multiuser} is determined as
\begin{equation}
C_{k_1,n}^* = 1, \quad k_1 = \arg\max\limits_k\; L_{k,n}^*, \quad 1\leq n \leq N,
\end{equation}
where the subcarrier is allocated to the worker with the largest value. Then, given the subcarrier-allocation scheme $\{C_{k,n}^*\}$, the latency-minimization problem is a special case of Problem (P1), whose solution can also be solved by Algorithm \ref{Ag:JointOptimalDesign}.

\section{Joint SUPPORT for DNN Models}\label{Section:DesignDNN}

In this section, DNN models are considered. Since Problem (P1) is not tractable in this case with the additional constraint (${\rm C_{dnn}}$), we propose an approximate solution method that leverages the result for decomposable model case, described as follows.% And the performance loss due to the approximation can be shown to be small. %based on the joint design of  SUPPORT in Section \ref{Section:DesignDecomposable}. However, 
%As mentioned, directly applying  the design in Section \ref{Section:DesignDecomposable} to implement DNN models may violate  the neuron allocation requirement and per-sample auxiliary matrix allocation requirement in Criterions \ref{Criterion:WStage} and \ref{Criterion:ZStage}. That says the additional parameter constraints (${\rm C_{dnn}}$) may not be satisfied.
%the parameters of the same subproblem, say the neuron problem defined in \eqref{Eq:WStep} for ${\bf W}$-stage or the problem of per-sample auxiliary matrix defined in \eqref{Eq:ZStep} for ${\bf Z}$-stage, may be allocated to different workers. 
%Thereby, additional communication among these workers is needed to solve the subproblem in one round. To yield the communication-efficient scheme, a practical method satisfying the requirements in Criterions \ref{Criterion:WStage} and \ref{Criterion:ZStage} are proposed as follows.
\begin{enumerate}
\item For both ${\bf W}$-stage and ${\bf Z}$-stage, solve the joint scheme of SUPPORT using the method in Section \ref{Section:DesignDecomposable} without considering Granularity Constraints \ref{Criterion:WStage} and \ref{Criterion:ZStage}. 

\item Given the subcarrier-allocation scheme, round the parameter allocation for each worker to satisfy Granularity Constraint \ref{Criterion:WStage} for ${\bf W}$-stage and  Granularity Constraint \ref{Criterion:ZStage} for  ${\bf Z}$-stage.
\end{enumerate}
The challenges lie in Step 2) and are two-fold. On one hand, how should the rounding indicator be designed to minimize the rounding loss. On the other hand, as each worker's number of parameters changes, the corresponding channel-capacity (or power) allocation and intra-worker parameter allocation among the assigned subcarriers should be redesigned. To tackle these challenges, in the sequel, we first propose a joint scheme of SUPPORT for DNN models. Then, the rounding scheme is designed accordingly and the resultant latency increase is analyzed.

\subsubsection{Joint SUPPORT for DNN Models}
Denote the solved one-round latency as $T^*$, the subcarrier-allocation policy as $\{C_{k,n}^*\}$, the spectrum efficiencies as $\{R_{k,n}^*\}$, the number of parameters of worker $k$ as $L_k^*$, the number of parameters uploaded by worker $k$ on subcarrier $n$ as $L_{k,n}^*$.

Consider an arbitrary worker, say worker $k$. If its number of parameters is rounded down to satisfy (${\rm C_{dnn}}$), the reduced number of parameters is denoted as $\Delta L_k^{\rm d}\geq 0$. If its number of parameters is rounded up, the additional number of parameters to be uploaded is denoted as $\Delta L_k^{\rm u}\geq 0$. Note that if worker $k$'s number of parameters is rounded down, no influence is caused to the one-round latency. Hence, only the case of being rounded up is considered in the sequel. Our aim is to design rounding scheme to minimize the resulted additional one-round latency.

Next, the joint scheme of SUPPORT is designed as
\begin{equation}\text{(Joint SUPPORT for DNN Models)}\label{Eq:PracticalDesignDNN}\quad
\left\{
\begin{aligned}
&C_{k,n} = C_{k,n}^*,\;R_{k,n} = R_{k,n}^*,\\
&\Delta L_{k,n} = L_{k,n}^* \times \dfrac{\Delta L_k^{\rm u}}{L_k^*},
\end{aligned}
\right.
\end{equation}
where $\Delta L_{k,n}$ is the number of additional parameters allocated to subcarrier $n$ for uploading, which is proportional to its currently uploaded number of parameters $L_{k,n}^*$. In \eqref{Eq:PracticalDesignDNN}, the allocation of subcarriers $\{C_{k,n}\}$ and the channel capacities $\{R_{k,n}\}$ of the assigned subcarriers remain the same. %And the number of additional parameters allocated to one subcarrier is proportional to its currently uploaded number of parameters, say $L_{k,n}^*$. 
%It can be easily verified that all constraints in Problem (P1) can be satisfied with the design in \eqref{Eq:PracticalDesignDNN} and the proof is omitted. 
Two concerns motivate us to design the joint SUPPORT scheme as \eqref{Eq:PracticalDesignDNN}. First,  the assigned subcarrier that can currently upload more updates of parameters can upload more additional parameters in the same additional latency. Second, the proportional additional parameter allocation together with the unchanged allocation of subcarriers and channel capacities can yield a simple upper bound of the additional latency for each worker, as shown in  the following lemma.
 %guarantee all constraints in Problem (P1) can be satisfied.
%With the practical design in \eqref{Eq:PracticalDesignDNN}, one upper bound of the additional latency for worker $k$ can be derived, as in the following lemma. 

\begin{lemma}[Additional Latency]\label{Lma:AddtionalLatency}
\emph{
Consider an arbitrary worker, say worker $k$, the design in \eqref{Eq:PracticalDesignDNN} results in an  upper bound of the minimum additional latency:
\begin{equation}\label{Eq:AdditonalLatency}
\Delta T_k \leq T^* \times \dfrac{\Delta L_k^{\rm u}}{L_k^*},
\end{equation}
where $T^*$ is the solved latency in Step 1), $\Delta T_k$, $\Delta L_k^{\rm u}$, and $L_k^*$ are the additional latency, the number of additional parameters after the rounding operation, and the solved number of parameters in Step 1) of worker $k$, respectively.
}
\end{lemma}
The proof of Lemma \ref{Lma:AddtionalLatency}  is straightforward and hence omitted.
%\proof See Appendix \ref{Apdx:LmaAddtionalLatency}.
Two observations can be made from Lemma \ref{Lma:AddtionalLatency}. On one hand, as mentioned in Example \ref{Rmk:SmallSizeSubproblem},  the size of the subproblems are far smaller than the problems of ${\bf W}$-stage and ${\bf Z}$-stage, i.e., $\Delta L_k^{\rm u} \ll L_k^*$. Therefore, \emph{the additional latency $\Delta T_k$ is small for all workers}. On the other hand, \emph{the round-up indicator, denoted as $I_k$, should be the ratio $I_k = \dfrac{\Delta L_k^{\rm u}}{L_k^*}$}.

\subsubsection{Parameter Rounding Scheme}
Note that Lemma \ref{Lma:AddtionalLatency} only gives the additional latency for one worker. To minimize the additional one-round latency, the rounding scheme is designed to make the workers with least $I_k$ to round up and the others to round down, described as follows.
\begin{enumerate}
\item Sort the round-up indicators $\{I_k\}$ from the least to the biggest and the new permutation is indexed by $k^{'}$, i.e., $\{I_{k^{'}}\}$ is sorted from the least to the largest.

\item Find the least $K_1^{'}$ following the new permutation $\{I_{k^{'}}\}$, which satisfies 
\begin{equation}\label{Eq:Round}
\sum\limits_{k^{'}=1}^{K_1^{'}} \Delta L_{k^{'}}^{\rm u} \geq \sum\limits_{k^{'}=K_1+1}^{K} \Delta L_{k^{'}}^{\rm d},
\end{equation}
where $L_{k^{'}}^{\rm u}$ is the additional number of parameters of worker $k^{'}$ when being rounded up and $\Delta L_{k^{'}}^{\rm d}$  is the reduced number of parameters when being rounded down. \eqref{Eq:Round} means that by rounding up $K_1^{'}$ workers with least round-up indicators, the parameters of all workers can satisfy Granularity Constraints \ref{Criterion:WStage} and \ref{Criterion:ZStage}.

\item The additional one-round latency is $\Delta T \leq T^* \times I_{K_1^{'}}$, where $T^*$ is the solved one round latency without considering Granularity Constraints \ref{Criterion:WStage} and \ref{Criterion:ZStage}.
\end{enumerate}

\section{Experimental Results}

\subsection{Experiment Setup}

The experimental settings are specified as follows unless specified otherwise. In the OFDM based PARTEL system, there are $K$ workers and $N$ subcarriers. The bandwidth of each subcarrier is $B=312.5$ kHz. The subcarrier channel gains $\{H_{k,n}\}$ are assumed to be i.i.d. Rayleigh fading with the average path loss of $10^{-3}$. The noise power density is set as $10^{-9}$ W/Hz. The workers' computation speeds $\{f_k\}$ are uniformly selected from the set $\{0.1,0.2,...,1.0\}\times 10^6$ parameters processed per second in one local computation iteration. The corresponding computation power factors $\{g_k\}$ are  uniformly selected from the set $\{0.1,0.2,...,1.0\}\times 10^{-16}$. The maximum power consumed by workers $\{P_k\}$ is set as $8$ W.  Both decomposable models and DNN models are trained using the PARTEL framework. The learning settings are as follows.
\begin{itemize}
\item \emph{Decomposable Model}: A $L_1$-regularized logistic regression task is considered, which trains a news-filtering model using the News20 dataset collected in \cite{lang1995newsweeder}. The model size is $1.24\times 10^6$. The training and test datasets have $15936$ and $3993$ samples respectively. $K=50$ workers with $N=80$ subcarriers are used to complete the task.

\item \emph{DNN Model}: The CNN model ``LeNet-5" proposed in \cite{lecun1998gradient} is trained on the MNIST dataset. In the ``LeNet-5" model, there are %2 convolutional layers, each followed by a pooling layer, and one fully connected layer and an output layer. There 
are $60,000$ parameters in total. The method of auxiliary variables in \cite{choromanska2019beyond} is used to train the ``LeNet-5" model at the PARTEL framework. In each training iteration, a mini batch of $50$ data samples is used. There are $469,400$ auxiliary variables in total. $K=30$ workers with $N=50$ subcarriers are used to complete the task. 
\end{itemize}

For comparison, three communication schemes are considered, described as follows.
\begin{itemize}
\item{Joint SUPPORT}: The joint schemes of SUPPORT proposed in Sections IV or V.

\item{Baseline}: The number of parameters computed by each worker is first allocated proportional to their computation capacity. Then, the subcarriers are allocated, which is a special case of the joint SUPPORT scheme.

\item{Greedy Scheme for FEEL}: The training samples are equally distributed among workers. Thereby, the computation latency and energy  of each worker is determined. The subcarrier allocation follows a greedy way. The subcarriers are randomly indexed and sequentially allocated from the $1$st to the $N$-th. The $i$-th subcarrier is allocated to the worker whose latency is currently the longest. Note that the latency minimization of one worker given the subcarrier allocation is simple and omitted.
\end{itemize}

\subsection{Decomposable Models}

\begin{figure}[t]
    \centering
    \subfigure[Training accuracy versus latency.]{\includegraphics[width=0.49\textwidth]{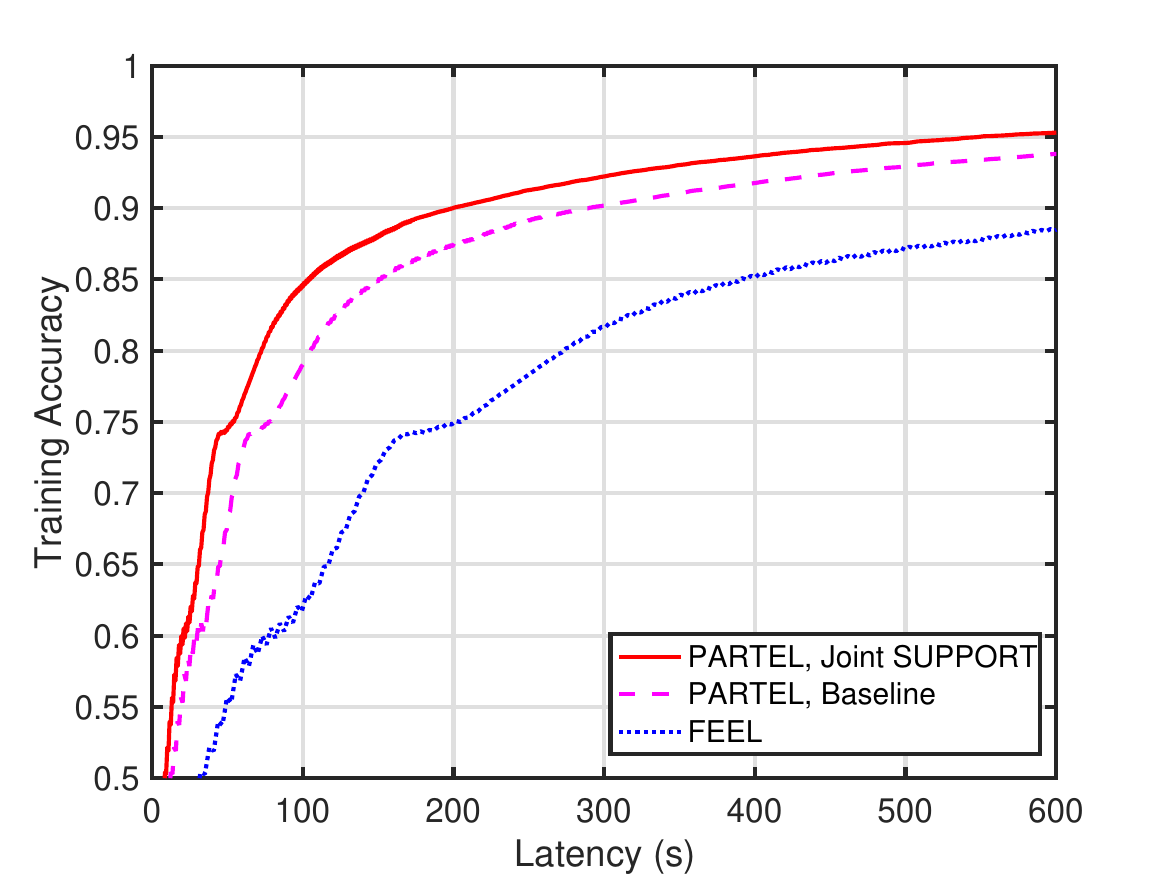}}
    \subfigure[Test accuracy versus latency.]{\includegraphics[width=0.49\textwidth]{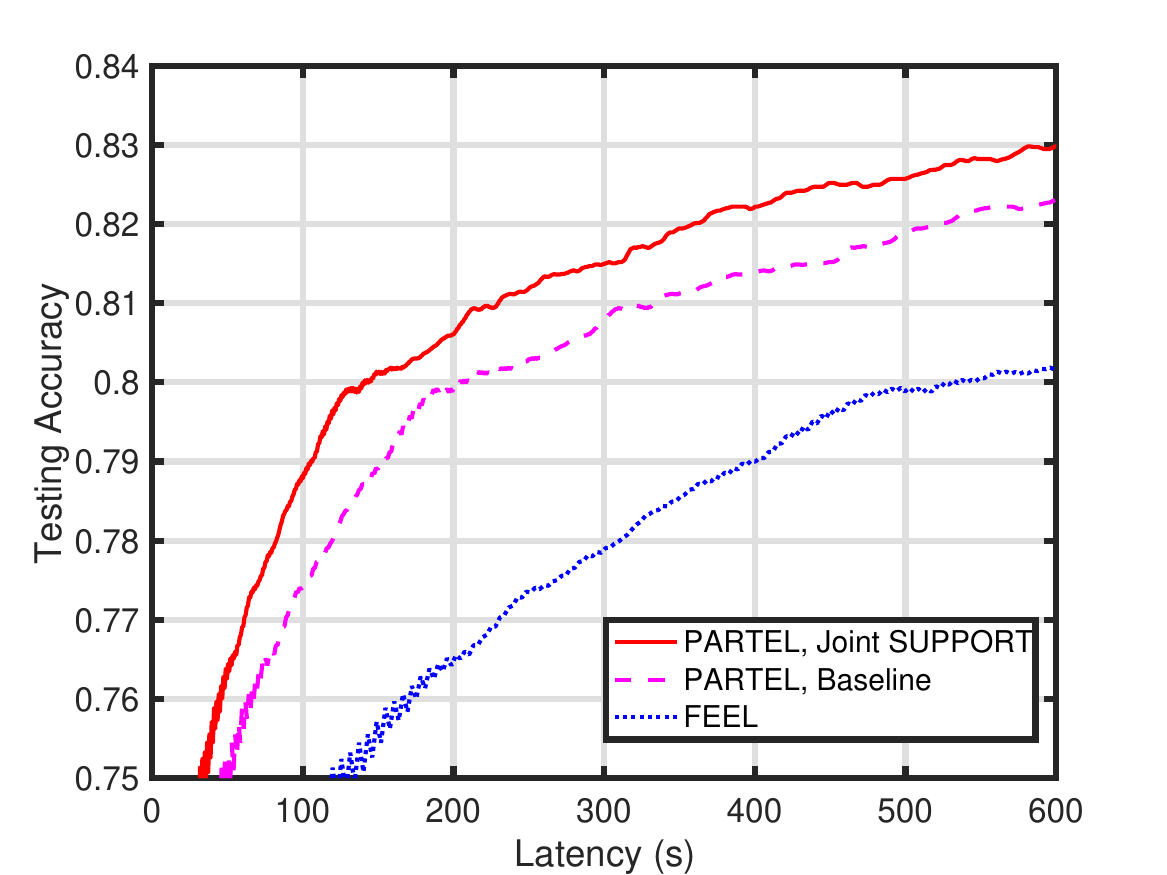}}
    \caption{Learning performance  versus (communication-plus-computation) latency.}\label{fig:LearningLR}    
\end{figure}

The learning performance of training the logistic regression model is compared in Fig. \ref{fig:LearningLR}. As observed, the model trained in PARTEL with the proposed joint SUPPORT converges much faster than the one  trained in FEEL with the greedy communication scheme, in which each worker uploads the updates of all parameters. Besides, the joint SUPPORT outperforms the baseline in terms of model convergence with a latency reduction of 31.06\% on average. That's because the allocations of parameters and subcarriers are sequentially designed in the baseline.

\begin{figure}[t]
    \centering
    \subfigure[Effect of Number of Workers.]{\includegraphics[width=0.49\textwidth]{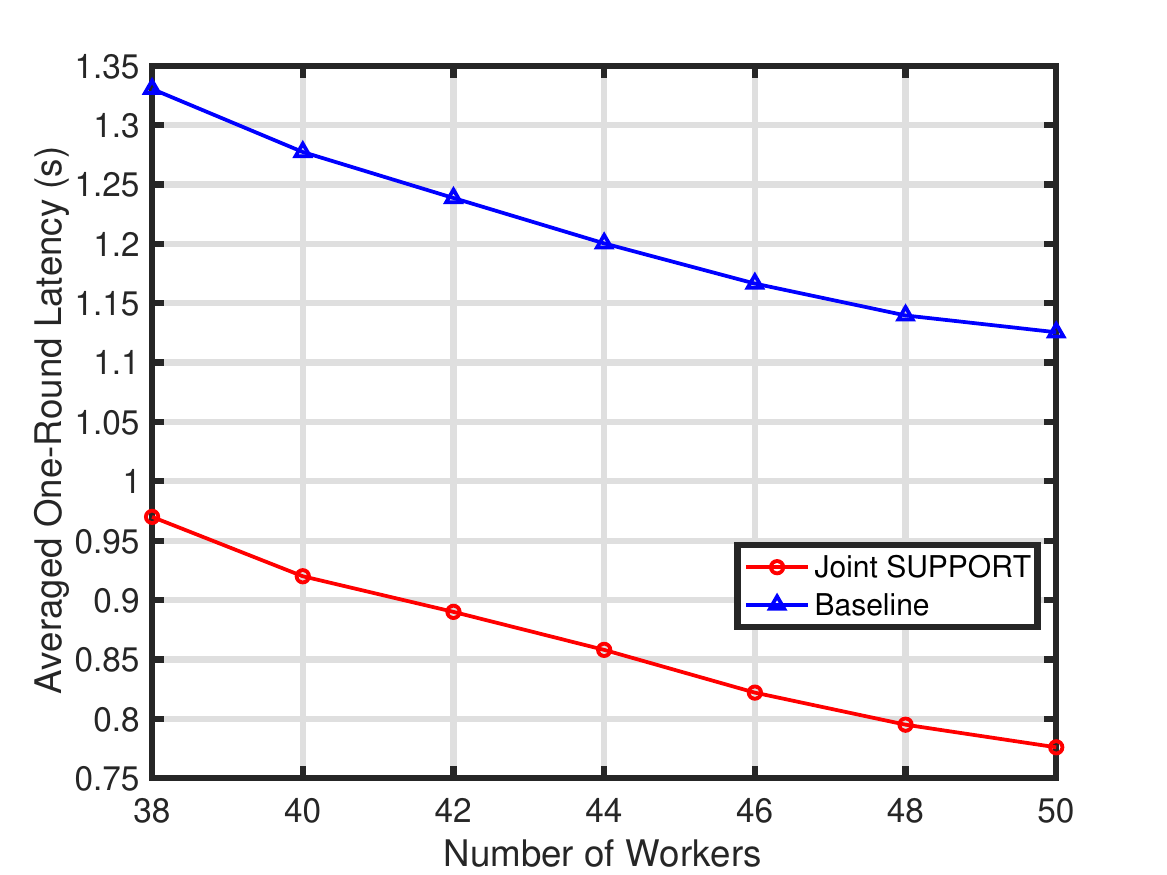}}
    \subfigure[Effect of Number of Subcarriers.]{\includegraphics[width=0.49\textwidth]{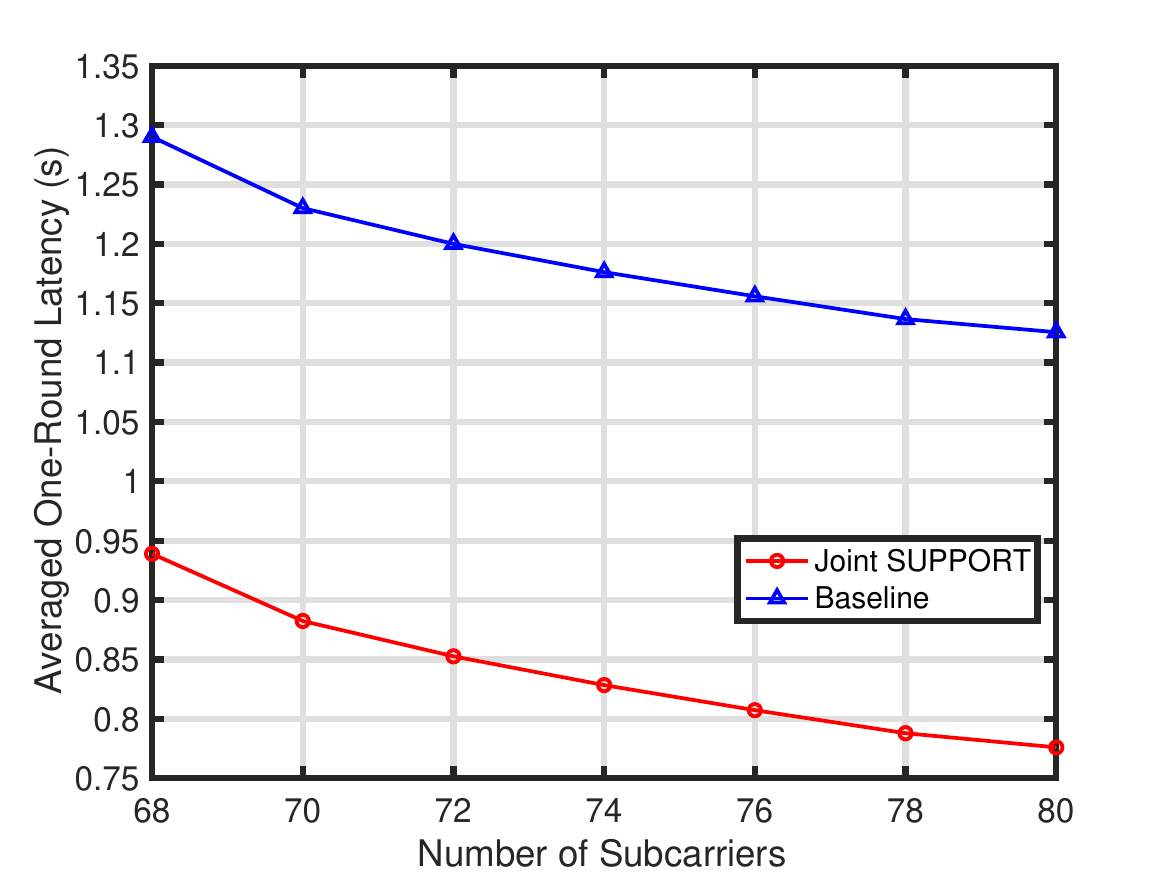}}
    \caption{Latency performance  versus (a) a varying number of workers and (b) subcarriers.}\label{fig:LatencyLR}
\end{figure}

Fig. \ref{fig:LatencyLR} shows the impacts of number of workers and subcarriers on the per-round latency. As observed, the per-round latencies of both schemes decrease as the number of workers or subcarriers increases. The reasons are as follows. More workers can provide more computation capacity and hence reduce the computation latency.  Moreover, more subcarriers allocated to workers can reduce the uploading latency.

\subsection{DNN Models}

\begin{figure}[t]
    \centering
    \subfigure[Training accuracy versus latency.]{\includegraphics[width=0.49\textwidth]{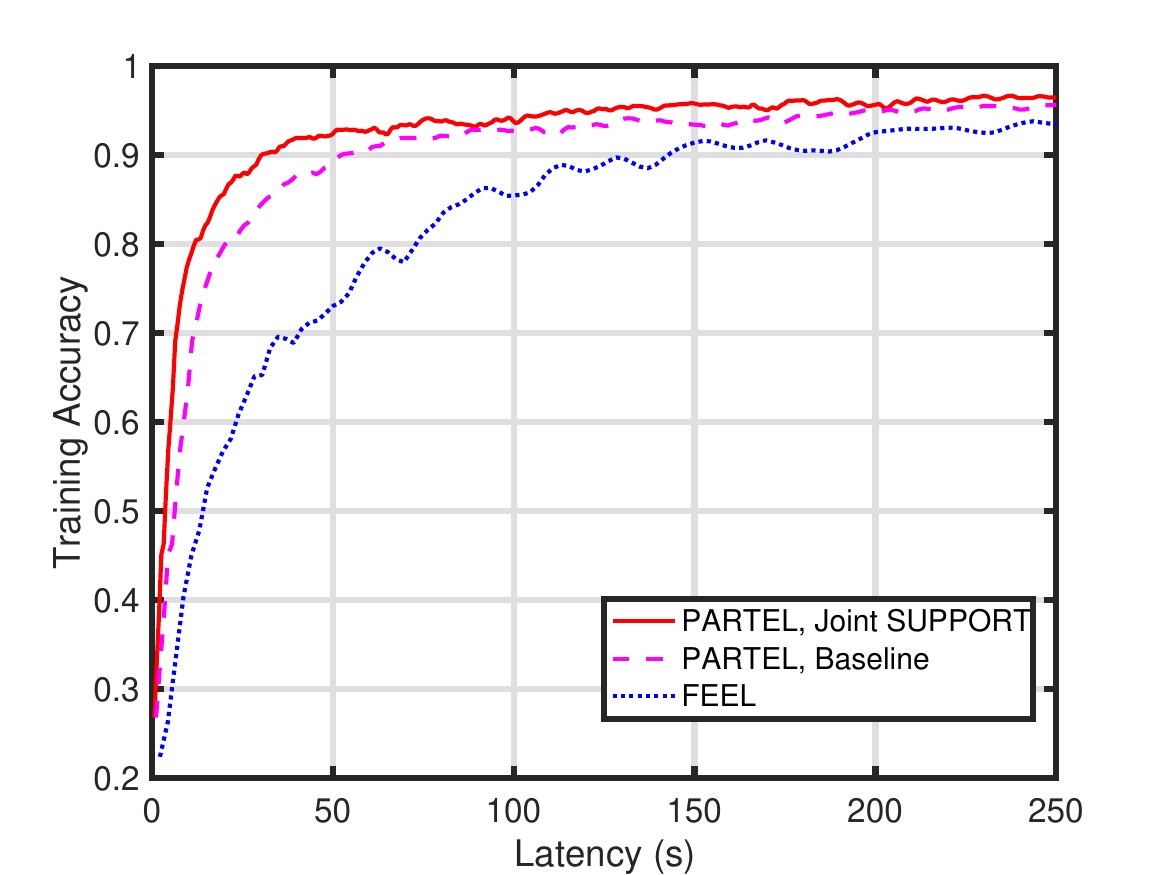}}
    \subfigure[Test accuracy versus latency.]{\includegraphics[width=0.49\textwidth]{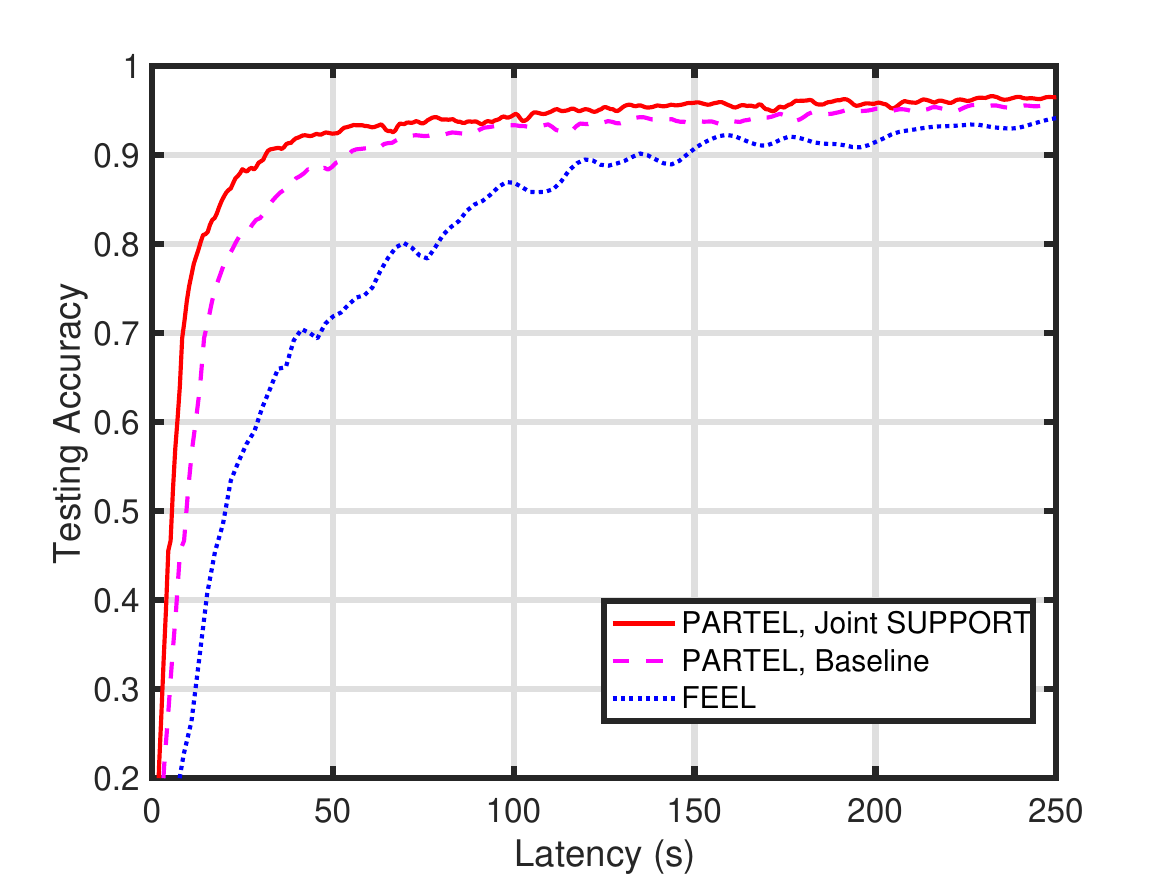}}
    \caption{Learning performance  versus (communication-plus-computation) latency.}\label{fig:LearningLeNet}    
\end{figure}

The learning performance of training the LeNet-5 is compared in Fig. \ref{fig:LearningLeNet}. In the figure, for FEEL, the optimizer is Adam (proposed in \cite{kingma2015adam}) and the corresponding learning rate is 0.002. For PARTEL, the optimizer is SGD. The learning rates for updating weights and auxiliary variables are 0.002 and 1, respectively. Although two stages (rounds) complete one training iteration in PARTEL using the joint scheme, it outperforms the FEEL using the greedy scheme in terms of  model convergence, as the latter has to upload the updates of all parameters in each round. Besides, the joint SUPPORT can reduce latency by 42.11\% compared to the baseline for the similar reason in the decomposable model case.

\begin{figure}[t]
    \centering
    \subfigure[Effect of Number of Workers.]{\includegraphics[width=0.49\textwidth]{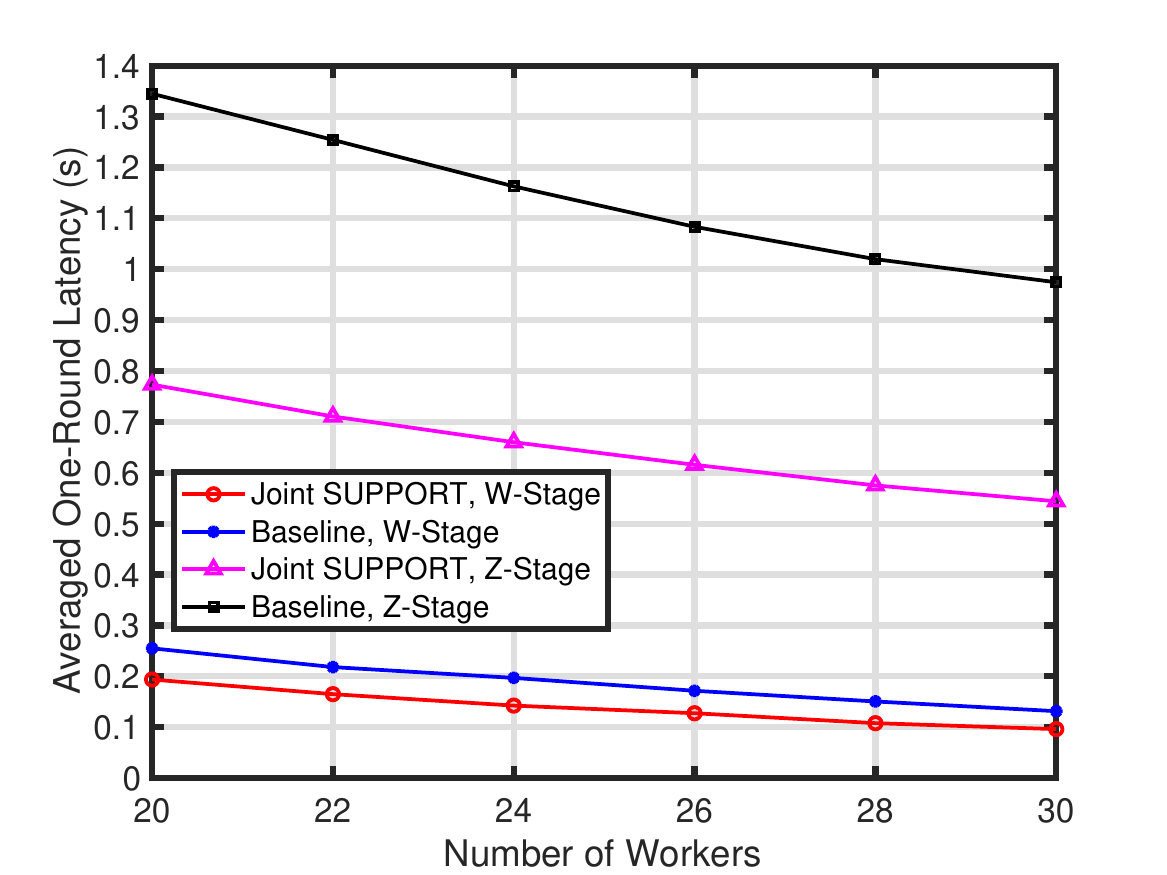}}
    \subfigure[Effect of Number of Subcarriers.]{\includegraphics[width=0.49\textwidth]{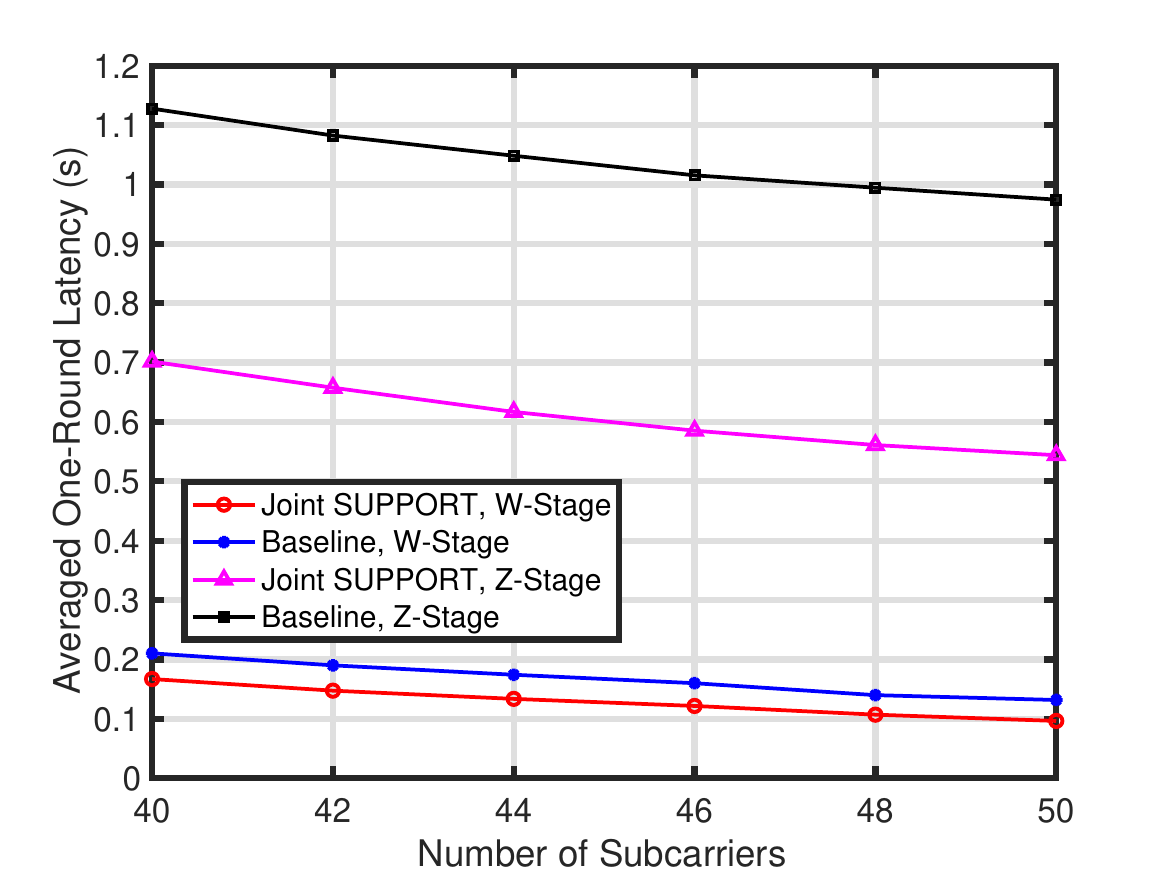}}
    \caption{Latency performance  versus (a) a varying number of workers and (b) subcarriers.}\label{fig:LatencyLeNet}
\end{figure}

The impacts of the number of workers and subcarriers on the latency performance of training LeNet-5 is compared in Fig. \ref{fig:LatencyLeNet}. As shown in the figure, the latencies of the two schemes for both ${\bf W}$-stage and ${\bf Z}$-stage decrease with the number of workers and subcarriers for the same reasons in the case of decomposable models.

The experimental results above show that our proposed joint scheme of SUPPORT has the best performance regarding learning latency and verifies our analysis.

\section{Concluding  Remarks}
In this paper, we have  presented a set of algorithms for jointly controlling parameter and sub-carrier allocation and power control, which can significantly reduce the latency of PARTEL deployed in a broadband system.  This work opens several  interesting directions. One is to take  the device  scheduling into consideration for further accelerating the training process.  In the context of  PARTEL, it is useful to jointly design the scheduler and parameter allocation.  Another interesting direction is to jointly control parameter allocation and local computation (e.g., processor speeds). In addition, the current joint design assuming OFDM can be extended to other advanced communication techniques such as non-orthogonal multi-access, massive MIMO, and over-the-air aggregation.

\appendix

%\subsection{Derivation of The Method of Auxiliary Variables}\label{Apdx:MAC}
%With the auxiliary variables, the problem in \eqref{Eq:DNN} can be equally derived as minimizing
%\begin{equation}\label{Eq:MAVP}
%\begin{aligned}
%\mathcal{L}({\bf W};{\bf Z}) &=  \sum\limits_{m=1}^M \left|  y_{m} -  f_{G+1} \left({\bf z}_{G,m};   {\bf W}_{G+1} \right)  \right|^2,\\
%\text{s.t. }\;  &{\bf z}_{g,m} = {\bf f}_g({\bf z}_{g-1,m}; {\bf W}_g), \quad 1<g\leq G,\; \& \; 1\leq m \leq M,\\
%&{\bf z}_{1,m} = {\bf f}_1({\bf x}_m; {\bf W}_g),\quad 1\leq m \leq M,
%\end{aligned}
%\end{equation}
%which, by using the quadratic-penalty method \cite{nocedal2006numerical}, can be further derived as the problem in \eqref{Eq:MAVP1}.

\subsection{Proof of Lemma \ref{Lma:EquivalentL}}\label{Apdx:LmaEquivalentL}

KKT conditions are used to show Lemma  \ref{Lma:EquivalentL}. The Lagrange function of Problem (P1) is in \eqref{Eq:apdxLagrange},
where $\{\mu_n\}$, $\lambda\geq0$, $\{\nu_k\geq0\}$, $\{\alpha_k\geq0\}$, and $\{\beta_{k,n}\geq0\}$ are multipliers. 

\begin{equation}\label{Eq:apdxLagrange}
\begin{aligned}
\mathcal{L} =& T + \sum\limits_{n=1}^N \mu_n \left(1 -  \sum\limits_{k=1}^K C_{k,n} \right) + \lambda \left(  L - \sum\limits_{k=1}^KL_k \right) + \sum\limits_{k=1}^K\nu_k \left( L_k - \sum\limits_{n=1}^N C_{k,n} L_{k,n} \right)\\
 + &\sum\limits_{k=1}^K\alpha_k \left(\dfrac{ E_k  } { T_k  } - P_k \right)  + \sum\limits_{k=1}^K\sum\limits_{n=1}^N\beta_{k,n}  C_{k,n} \left( T^{\rm cmp}_k +  T^{\rm com}_{k,n}   - T \right),
\end{aligned}
\end{equation}
Then, consider an arbitrary subcarrier-allocation scheme $\{C_{k,n}\}$, KKT conditions are necessary to solve the problem. Some related KKT conditions are given below:
\begin{equation}\label{Eq:apdxKKT1}
\left\{
\begin{aligned}
& \dfrac{\partial \mathcal{L} }{ \partial T } = 1-C_{k,n}\beta_{k,n} = 0,\quad 1\leq k \leq K,\\
&  \beta_{k,n}  C_{k,n} \left( T^{\rm cmp}_k +  T^{\rm com}_{k,n}  - T \right) = 0,\quad \forall (k,n),\\
%&\beta_{k,n}\geq 0,\quad \forall (k,n).
\end{aligned}
\right.
\end{equation}
From the first condition in \eqref{Eq:apdxKKT1}, we can show that 
%\begin{equation}
$\{\beta_{k,n} \neq 0, \; \forall C_{k,n}=1\}$,
%\end{equation}
which, together with the second condition in \eqref{Eq:apdxKKT1}, can show that
%\begin{equation}\label{Eq:apdxEqL}
 $\{T^{\rm cmp}_k +  T^{\rm com}_{k,n} = T,  \; \forall C_{k,n}=1\}$.
%\end{equation}
Note that the above condition is necessary for arbitrary subcarrier-allocation schemes. Hence, it is a necessary condition to solve (P1).

\subsection{Proof of Lemma \ref{Lma:RelationModelSizeLatency} }\label{Apdx:LmaRelationModelSizeLatency}

First, we show that the equality of the third and forth constraints in Problem (P3) should be achieved. The Lagrange function of (P3) is
\begin{equation}
\begin{aligned}
\mathcal{L} = &\sum\limits_{k=1}^K L_k + \sum\limits_{n=1}^N \mu_n \left(\sum\limits_{k=1}^K C_{k,n} - 1 \right)  + \sum\limits \lambda_k \left[ L_k -   \sum\limits_{n=1}^N \dfrac{C_{k,n} R_{k,n}  }{  \tau }\left( T -   \dfrac{ L_k }{f_k} \right) \right] \\
+& \sum\limits \nu_k \left(E_k  - P_k T \right),
\end{aligned}
\end{equation}
where $\{\mu_n\}$, $\{\lambda_k\geq 0\}$, and $\{\nu_k\geq 0\}$ are multipliers. Using KKT conditions and the similar approaches in Appendix \ref{Apdx:LmaEquivalentL}, we can show
% are necessary to achieve the optimal solution, some of which are given as
%\begin{equation}
%\left\{
%\begin{aligned}
%&\dfrac{\partial \mathcal{L}  }{ \partial L_k} = 0, \; 1\leq k \leq K,\;  \; \dfrac{\partial \mathcal{L}  }{ \partial R_{k,n} } = 0,\; \forall (k,n),\\
%%&\dfrac{\partial \mathcal{L}  }{ \partial R_{k,n} } = 0,\quad \forall (k,n),\\
%&  \lambda_k \left[ L_k -   \sum\limits_{n=1}^N \dfrac{C_{k,n} R_{k,n}  }{  \tau }\left( T -  \dfrac{ L_k }{f_k} \right) \right] = 0, \quad 1\leq k \leq K,\\
%&\nu_k \left(E_k - P_k T \right) = 0,\quad  1\leq k \leq K,\\
%%&\lambda_k,\nu_k \geq0, \quad 1\leq k \leq K,
%\end{aligned}
%\right.
%\end{equation}
%which shows 
that $\{\lambda_k\neq 0,\; 1\leq k \leq K\}$ and $\{\nu_k\neq 0\; 1\leq k \leq K\}$ and the equalities of the third and forth constraints in Problem (P3) should be achieved.

Then, consider $T_1<T_2$. When $T=T_1$, denote the optimal solution of (P3) as $\{C_{k,n,1}^*\}$, $\{L_{k,1}^*\}$, $\{R_{k,n,1}^*\}$, and the maximum updatable model size as $L^*(T_1)$. 

Next, let $T=T_2$, $\{C_{k,n,2} = C_{k,n,1}^*\}$, and $\{R_{k,n,2} = R_{k,n,1}^*\}$. By substituting $L_{k,1} = L_{k,1}^*$ into the third and forth conditions in Problem (P3), the equalities are not achieved. This shows that the updatable number of parameters by each worker, denoted as $\{L_{k,2}\}$, can be larger, i.e.,
%\begin{equation}
$L_{k,2}>L_{k,1}^*$.
%\end{equation}
It follows that
%\begin{equation}
$\sum\nolimits_{k=1}^K L_{k,2}>\sum\nolimits_{k=1}^K L_{k,1}^* = L^*(T_1)$.
%\end{equation}
Furthermore, the optimal solution for $T=T_2$ satisfies $L^*(T_2)\geq  \sum\nolimits_{k=1}^K L_{k,2}$. Hence, we have 
%\begin{equation}
$L^*(T_2) > L^*(T_1)$.
%\end{equation}

\subsection{Proof of Lemma \ref{Lma:Convexity}}\label{Apdx:LmaConvexity}

%First, the two variable transforms in \eqref{Eq:VariableTransform} can be equivalently derived as
%\begin{equation}\label{Eq:ApdxVariableTransform}
%\left\{
%\begin{aligned}
%&L_k = f_k \left( t - \dfrac{ 1 }{ \phi_k } \right), \\
%&R_{k,n} = \dfrac{ \tilde{R}_{k,n} }{ C_{k,n} }.
%\end{aligned}
%\right.
%\end{equation}
First, the third constraint in Problem (P3), by dividing $\left( T -   L_k /f_k \right)$ on both sides and substituting the variable transformations in \eqref{Eq:VariableTransform}, can be derived as the third constraint in Problem (P4):
%\begin{equation}
%\sum\limits_{n=1}^N \dfrac{C_{k,n} R_{k,n}  }{  \tau } \geq L_k \left( T -    \dfrac{ L_k }{f_k} \right)^{-1},\quad 1\leq k \leq K,
%\end{equation}
%which, by substituting the variable transformations in \eqref{Eq:VariableTransform}, can be derived as the third constraints in Problem (P4):
%\begin{equation}
$\left\{\sum\nolimits_{n=1}^N   \tilde{R}_{k,n} /  \tau \geq  f_k \left( T \phi_k - 1 \right),\; 1\leq k\leq K\right\}$.
%\end{equation}
Obviously, the feasible region of the above constraint is a convex set.
Then, by substituting $E_k  $ in \eqref{Eq:WorkerE1}, dividing $\left( T -   \dfrac{ L_k }{f_k} \right)$ on both sides, and substituting the variable transformations in \eqref{Eq:VariableTransform}, the forth constraint in Problem (P3) can be equally derived as 
%\begin{equation}
 %\sum\limits_{n=1}^N  \dfrac{ C_{k,n} \big( 2^{ \frac{ \tilde{R}_{k,n} } { BC_{k,n} } }-1 \big)  \sigma^2 }{ h_{k,n} }  + g_kf_k^2L_k \left( T -   \dfrac{ L_k }{f_k} \right)^{-1} \leq (P_kT-\xi) \left( T -   \dfrac{ L_k }{f_k} \right)^{-1}, \; 1\leq k \leq K,
%\end{equation}
%which, by substituting the variable transformations in \eqref{Eq:VariableTransform}, can be derived as 
the forth constraint in Problem (P4):
\begin{equation}\label{Eq:ApdxPowerCons}
\sum\limits_{n=1}^N  \dfrac{C_{k,n}  \sigma^2 \big( 2^{ \frac{\tilde{R}_{k,n} }{BC_{k,n} } }-1 \big) }{ h_{k,n} }   +  g_kf_k^3 \left( \phi_kT- 1 \right) \leq ( P_kT - \xi) \phi_k,\; 1\leq k\leq K.
\end{equation}
In \eqref{Eq:ApdxPowerCons}, the first term is a convex function as $f(x,y) = x e^{y/x} $ is convex. Thereby, the feasible region of the constraint in \eqref{Eq:ApdxPowerCons} is a convex set.
Besides, the objective function and other constraints are convex. Thus, Problem (P4) is convex.

\bibliographystyle{ieeetr}
\bibliography{reference}

\end{document}